\documentclass[12pt]{iopart}
\pdfoutput=1
\usepackage{amsfonts}
\usepackage{graphicx}
\usepackage{iopams}
\usepackage[mathscr]{euscript}
\usepackage{ragged2e} 

\usepackage{xcolor}
\usepackage{bm}
\usepackage{subfigure}
\usepackage{textcomp}
\usepackage{amsmath}
\usepackage{cite}
\usepackage{float}
\usepackage{soul}
\usepackage{stackrel}
\usepackage{ulem}
\usepackage{multirow}
\usepackage{MnSymbol}

\usepackage{stackengine}

\usepackage{booktabs}

\usepackage{tikz}
\usepackage{esint}
\usetikzlibrary{arrows,arrows.meta}
\newcommand\semiInt[1][1]{%
    \begin{tikzpicture}[scale=0.2]
        \coordinate (center) at (0.2,0.55);
        \draw[black,-{>[scale=0.6]}] (0, 0.0) + (center) arc (180:0:0.5);
       $\int$
    \end{tikzpicture}
    }

\usepackage[colorlinks,linkcolor=blue,urlcolor=blue,citecolor=blue]{hyperref}
\allowdisplaybreaks

\pdfoutput=1
\usepackage{esint}

\usepackage{hyperref}

\begin{document}

\title{Nonequilibrium fluctuations in a harmonic trap with annealed stochastic stiffness}

\author{Deepak Gupta}
\ead{phydeepak.gupta@gmail.com}
\address{Institut für Physik und Astronomie, Technische Universität Berlin, Hardenbergstraße 36, D-10623 Berlin, Germany}

\author{Sabine H. L. Klapp}
\address{Institut für Physik und Astronomie, Technische Universität Berlin, Hardenbergstraße 36, D-10623 Berlin, Germany}

\vspace{10pt}

\begin{abstract} 
We provide a comprehensive analysis of the positional dynamics and average thermodynamics of an overdamped Brownian particle subject to both, harmonic confinement and annealed disorder due to a temporarily fluctuating trap stiffness.
We model this stiffness via a stationary Ornstein-Uhlenbeck (OU) process whose correlation time can be tuned from white noise to a quenched limit. We analytically calculate the positional distribution in these limits and provide exact expressions for the 
$n$th positional moments at finite correlation times, revealing important insights regarding stationarity. Further, 
we analyze the average work performed on the particle and the heat dissipated into the environment at all times, illustrating the nonequilibrium character of the system and its relaxation into a steady state. Our analytical results are validated by numerical Langevin simulations.
\end{abstract}


\section{Introduction}
\label{intro}
It is well established that small-scale systems are typically subject to strong temporal fluctuations~\cite{Seifert_2012-ST,shiraishi2023introduction}. Well-known examples are colloids, polymers, or bio-molecules surrounded by a solvent (the bath) where the typical energy scales of the embedded ``system'' are comparable to the thermal energy, $k_{\rm B}T$ (with $k_{\rm B}$ denoting Boltzmann’s constant and $T$ the temperature) provided by the environment. Assuming a spatially homogeneous environment, and a perfect separation of time scales between system and bath, these thermal fluctuations are often taken account in the spirit of Brownian motion, involving a (coarse-grained)
fluctuating force that is Gaussian-distributed,
and temporarily uncorrelated~\cite{van1992stochastic}.

Recently, significant attention has been devoted to systems that are subject to several, not only thermal, types of fluctuations or random disorders.
Prominent examples are systems in heterogeneous environments where the bath is not just a homogeneous background (with spatially constant viscosity and diffusion constant). These play a crucial role in various biological and physical contexts
\cite{expt_1, METZLER20001,disorder-1,hetro-1,hetro-2}, such as in intracellular transport~\cite{disorder-2}, protein dynamics on fluctuating membranes~\cite{Seifert-prot-1,Seifert-prot-2}, motion of RNA-protein complexes in live Escherichia coli and Saccharomyces cerevisiae cells~\cite{LAMPO2017532}, as well as in living yeast cells~\cite{yeast-cells}, to name a few.

When investigating disorder effects in theoretical approaches, one has to distinguish between quenched and annealed disorder. Quenched disorder implies that some relevant random parameters do not evolve in time (which implies that the usual ensemble averages have to be supplemented by a disorder average). Exemplary works on the impact of quenched disorder on systems subject to thermal fluctuations concern, e.g.,
the flocking of active particles~\cite{quench-3}, phase transitions in random-field Ising model~\cite{qunech-1} and spin glasses~\cite{quench-2}, transport in rocking ratchets~\cite{ratchets-1}, but also ecological communities~\cite{quench-4,population-quench,galla2024generatingfunctionalanalysisrandomlotkavolterra}, pattern-forming systems~\cite{Yizhaq_2016}, and relaxation processes~\cite{Jan-M-1}.
In contrast, annealed (or dynamic) disorder means that the parameters characterizing the disorder change in time and are, therefore, in the same time-scales as that of the dynamics of the system. This implies that the usual time-scale separation cannot be performed.
Annealed (dynamic) disorder has gained significant attention in models involving fluctuating diffusivity~\cite{DF-1,DF-2,DF-3,Jain2017,SNGSdiffusion,Santra_2022, MK-SK}, consumer-resource dynamics~\cite{zanchetta2025emergenceecologicalstructurespecies}, Lotka--Volterra systems~\cite{LVA-1}, and linearly interacting particles' model~\cite{Ferraro_2025}.

Here, we briefly review the previous research on the annealed disorder.
Besides fluctuating diffusion coefficients~\cite{DF-1,DF-2,DF-3,Jain2017,SNGSdiffusion,Santra_2022, MK-SK}, annealed disorder
occurs, e.g., when the surrounding medium is near a critical point and exhibits strong density fluctuations, generating a fluctuating viscosity~\cite{ROZENFELD1998409,Huang2020}. Similarly, stochastic variations in particle mass have motivated models of Brownian particles with short-range attractive interactions and dynamically forming and dissociating clusters~\cite{gitterman2012oscillator-1,random_mass_1,Huang2020}. 
Another variant are stochastic modulations of the underlying potential~\cite{Alston_2022,Gomez-Solano_2010, Apal-w-1,HaenggiBartussek1996, DOERING19981}, and by dragging a probe coupled to fluctuating correlated field~\cite{Venturelli_2024}. Moreover, stochastic modulation of potentials has also gained interest for implementing finite-time stochastic resetting protocols~\cite{reset_linear,SRSR-pal,goerlich2024tamingmaxwellsdemonexperimental,Besga_2020} and analyzing the associated thermodynamic costs~\cite{olsen2024thermodynamic,gupta2022work, olsen2024thermodynamic2, gupta2025thermodynamiccostrecurrenterasure}.
Along the same line, researchers have discussed the Brownian particle's position fluctuations by stochastically switching on-off a $V$-shape potential~\cite{IRP_1}, and a harmonic potential~\cite{IRP-2}. Ref.~\cite{IRP-3} discusses the correlations of many particles systems in a harmonic trap with stiffness switching between two values. Further, Refs.~\cite{Shuttling-1,Shuttling-2} discusses the shuttling of a quantum particle by stochastically modulating the stiffness and minimum of the trap.

In the present work, we focus on an overdamped Brownian particle in a fluctuating harmonic potential, i.e., a fluctuating trap.
Previous works have addressed the positional dynamics in case of a fluctuating trap position (modeled as an Ornstein-Uhlenbeck noise~\cite{Gomez-Solano_2010, Apal-w-1}). Further, Ref.~\cite{Alston_2022} discusses the stationary state entropy production of a system in fluctuating harmonic potentials. 
However, a comprehensive understanding of the statistical and thermodynamic properties of a Brownian oscillator with simultaneously or independently fluctuating trap minimum and trap stiffness for a general stochastic fluctuating rule remains limited.

Here, we aim for a detailed analysis of the transient and steady-state dynamics of a harmonically confined Brownian particle in presence of a 
fluctuating stiffness described as 
an Ornstein-Uhlenbeck process with persistence time $t_p$. We also discuss the thermodynamic energy flows. The model may mimic, e.g., a system in a fluctuating potential energy landscape as in occurs in protein folding dynamics~\cite{protein} where pressure fluctuations change the stiffness of the local harmonic potential (see Fig.~2 in Ref.~\cite{protein}) without altering the barrier heights. (See also Ref.~\cite{Alston_2022} and references therein for more motivation on this topic.)
In our analysis we first obtain the positional distribution in the absence of thermal bath. In the presence of bath, we exactly compute the position distributions in the limit of vanishing ($t_p\to 0$, white noise limit) and long persistent time ($t_p\to \infty$, quenched noise limit). For general $t_p$, we employ the method of subordination~\cite{MOS-1,MOS-2} to exactly compute the $n$th position's moment. We also discuss the generalization to arbitrary stiffness fluctuations in the limit of weak stiffness noise. Further, to understand the energy cost associated with the stochastic switching of stiffness, we compute exactly the average work performed on the particle and average heat exchanged by the particle with the heat bath at arbitrary times $t$.

Given that a similar (yet not exactly the same) system has been studied before~\cite{Cocconi_2024}, we briefly summarize the main differences: First, 
we employ a different white noise's coefficient (motivated from Ref.~\cite{zanchetta2025emergenceecologicalstructurespecies}) in the stiffness-equation~\eqref{k-eqn-2}; this enables us to explore various noise regimes going from white to quenched disorder in stiffness, 
when the persistence time becomes large.
Second, for this quenched noise limit, we also provide
the exact positions' distribution. Third, the positions' fluctuations for general persistent time $t_p$~\eqref{k-eqn-2} are obtained using the method of subordination~\cite{MOS-1,MOS-2} for non-zero initial condition (which is important for the short-time behavior), whereas Ref.~\cite{Cocconi_2024} uses another method with zero initial condition. Finally, our calculations provide the average heat flow and work done not just in the stationary state~\cite{Cocconi_2024}, but at all times. Overall, our calculations complement the calculations in~\cite{Cocconi_2024} and provide different technical routes.

The rest of the paper is organized as follows. In the next section~\ref{sec:model}, we present the model. Section~\ref{sec:Deq0} discusses the probability density function of the position of the particle and its associated large deviation function in the absence of thermal noise ($T=0$). Then, we discuss the position distribution for the case when stiffness behaves as a white noise (i.e., $t_p\to 0$) in Subsec.~\ref{sec:WN-main}. Subsection~\ref{sec:QN-main} presents the calculations to obtain the position distribution for stiffness behaving as Gaussian quenched noise (i.e., $t_p\to \infty$). For finite non-zero $t_p$, we exactly compute the $n$th position moment and discuss their stationarity in the long-time limit in Subsec.~\ref{sec:gen_tp}. Then, we discuss the average work and average heat flow in Sec.~\ref{sec:thermo}. We summarize our main findings in Sec.~\ref{sec:summ}. \ref{sec:FP-WN} shows the derivation of the Fokker-Planck equation in the white noise limit. \ref{sec:I1I2} discusses the solutions of two integrals. \ref{app:work} discusses the detailed calculations to compute the average work.

\section{Model}
\label{sec:model}
We consider a Brownian particle coupled to a heat bath of temperature $T$, in a one-dimensional harmonic trap. The trap's net stiffness $[\sigma_0\kappa + \sigma k(t)]$, consists of two contributions, and is stochastic in time.
The first one, $\sigma_0 \kappa$ is independent of time (where $\kappa>0$), {\color{black}and represents} the static stiffness. Moreover, the time-independent parameter $\sigma_0$ can be either $+1$ or  $-1$ or $0$ depending on the situation if the static harmonic trap (characterized by $\kappa$) is either present [bounded ($\sigma_0 = 1$) or unbounded ($\sigma_0 = -1$)] or absent ($\sigma_0=0$). The second contribution involves the stochastic stiffness $\sigma k(t)$, where the time-independent parameter $\sigma$ controls the strength of the actual stochastic trap' stiffness $k(t)$.  
We specifically model the stochasticity of the additional 
stiffness $k(t)$ by a colored noise [see Eq.~\eqref{k-eqn} below]. [Later in Sec.~\ref{sec:gen_tp}, we will discuss a case of general $k(t)$.] 
Thus, the following Langevin equations of motion describe the full system:
\begin{subequations}
\label{eqns}
\begin{align}
    \dot x &= - [\sigma_0 t_k^{-1} + \sigma \gamma^{-1}k(t)]x(t) + \sqrt{2D}\eta(t) \label{x-eqn}\ ,\\
    \dot k &= -\dfrac{k(t)}{t_p} + \sqrt{A\dfrac{\theta t_k+2t_p}{t_p^2}}\xi(t)\label{k-eqn} \ ,
\end{align}
\end{subequations}
where the dot denotes a time derivative, and  $t_k\equiv \gamma/\kappa$ is the harmonic trap's relaxation time in the absence of $k(t)$ 
with friction constant $\gamma$. Further, $D=k_{\rm B}T/\gamma$ is the diffusion constant, and $\eta(t)$ and $\xi(t)$, respectively, are Gaussian white noises with zero mean and delta correlation in time, i.e., $\langle \eta(t)\eta(t')\rangle =\delta(t-t')$ and $\langle \xi(t)\xi(t')\rangle =\delta(t-t')$. Also, $\eta(t)$ and $\xi(t)$ are independent of each other, i.e., $\langle \eta(t)\xi(t')\rangle =0$ for all $t,t'$. Here, the 
brackets $\langle \cdots \rangle$ indicate the ensemble average over the noise realizations.

In Eq.~\eqref{k-eqn}, $t_p$ is the persistence time. The prefactor $\sqrt{A(\theta t_k + 2t_p)/t_p^2}$ of the noise $\xi(t)$ 
is motivated from Ref.~\cite{zanchetta2025emergenceecologicalstructurespecies} where it was introduced to understand the role of environmental fluctuations on the species abundance distribution in an ecological community. Later in this section, we will discuss
the significance of this prefactor in the present context. We also note that the prefactor is different from that {\color{black}chosen} in Ref.~\cite{Cocconi_2024}.

{\color{black}In Eq.~\eqref{eqns}, we rescale time, the spring constant, and the parameter $A$ using the relaxation time $t_k$, and the static spring constant $\kappa$, respectively, $t/t_k\to t$, $t_p/t_k\to t_p$, $k/\kappa \to k$, $A/\kappa^2\to A$.}
The Langevin equations~\eqref{eqns} then read
\begin{subequations}
\label{eqns-21}
\begin{align}
    \dot x &= - [\sigma_0 + \sigma k(t)]x(t) + \sqrt{2Dt_k}\eta(t) \label{x-eqn-21}\ , \\ 
        \dot k &= -\dfrac{k(t)}{t_p} + \sqrt{A\dfrac{\theta+2t_p}{t_p^2}}\xi(t) \label{k-eqn-21} \ .
\end{align}
\end{subequations}

The 
probability density function of $x$ depends on the fluctuations of $\eta(t)$ and $k(t)$. 
Since the distribution of $\xi(t)$ is Gaussian, the probability density function of stiffness $k(t)$ at each time $t$ is also Gaussian, with mean $\mu_k$ and variance $\Sigma_k^2$
\begin{subequations}
\label{k-mom}
  \begin{align}
\mu_k &\equiv \langle{k(t)}\rangle = k_0 e^{-t/t_p} \label{mean-k}\ , \\
    \Sigma_k^2&\equiv \langle{\delta k(t)^2} \rangle  
     = A\dfrac{\theta+2t_p}{2t_p}(1 - e^{-2t/t_p}) \ ,\label{var-k}
\end{align}  
\end{subequations}
for fixed initial condition $k_0 \equiv k(0)$. Here, $\delta k(t)\equiv k(t) - \langle {k(t)}\rangle$ is the deviation of $k$ from its mean value. Further, it is straightforward to 
compute the two-point correlation function as
\begin{align}
    \langle{\delta k(t)\delta k(t')} \rangle = A\dfrac{\theta+2t_p}{2t_p}[e^{-|t-t'|/t_p} - e^{-(t+t')/t_p}] \label{twopoint-1}\ .
\end{align}
In the long-time limit (i.e., $t\gg t_p$), 
the distribution of $k(t)$
becomes stationary,
and its explicit expression is given by
\begin{align}
    p_{\rm ss}(k)= \dfrac{1}{\sqrt{2\pi \Sigma_{k,\rm ss}^2}} \exp\bigg[-\dfrac{(k-\mu_{k,\rm ss})^2}{2\Sigma_{k,\rm ss}^2}\bigg]\ ,\label{pk-ss}
\end{align}
with mean $\mu_{k,\rm ss}=0$, and variance $\Sigma_{k,\rm ss}^2 = A\frac{\theta+2t_p}{2t_p}$. 
Further, the two-point correlation~\eqref{twopoint-1} becomes 
\begin{align}
    \langle{\delta k(t)\delta k(t')} \rangle_{\rm ss}= A\dfrac{\theta+2t_p}{2t_p}e^{-|t-t'|/t_p} \label{corr-k} \ .
\end{align}
Henceforth, we drop $\delta$ from $\delta k$ as the mean of $k$ is zero in the stationary state. 
In the {\color{black}limiting} cases of zero and diverging $t_p$, the two point correlation function behaves as 
\begin{align}
    \langle{k(t) k(t')} \rangle_{\rm ss}= \begin{cases} A\theta\delta (t -t')\qquad &
    \qquad  t_p\to 0 \\
    A\qquad  &
    \qquad  t_p\to \infty
    \label{corr-k-lim} 
    \end{cases}\ .
\end{align}
Equation~\eqref{corr-k-lim} reveals that the choice of the prefactor in front of $\xi(t)$~\eqref{k-eqn-21} (considered previously in Ref.~\cite{zanchetta2025emergenceecologicalstructurespecies}) allows us explore different types of $k(t)$. 
These are Gaussian white noise ($t_p\to 0$), 
Gaussian quenched noise ($t_p\to \infty$),  
and colored noise for finite non-zero $t_p$.
The Gaussian quenched noise ($t_p\to \infty$) implies that the stiffness's value $k(t)$ is fixed for all time \eqref{corr-k-lim}, but 
is drawn initially from a Gaussian distribution with zero mean and $A$ variance. 
Thus, the stochasticity of the additional noise $k(t)$ arises from its initial condition. 

In contrast to the stiffness's distribution $p(k,t)$, the probability density function $P(x,t|x_0)$ is not expected to be Gaussian, for the particle starting from $x_0$. In the following, we aim to compute the full positions' fluctuations 
of the particle with initial condition $x_0$ and $k_0$, where the latter 
{\color{black}is} drawn from the stationary distribution~\eqref{pk-ss}.

\section{Absence of heat bath: $T=0$}
\label{sec:Deq0}
{\color{black}In this section, we 
focus on a harmonic oscillator in the absence of external heat bath, i.e., $D=0$.  
Therefore, the system is described by the Langevin equations~\eqref{eqns-21} in the absence of $\eta(t)$.
Substituting $x = e^y$ (for $x\geq 0$) in Eq.~\eqref{x-eqn-21}, we have}
\begin{align}
   \dfrac{dy(t)}{dt} = -[\sigma_0 + \sigma k(t)]\ , \label{y-eq}
\end{align}
where $y\in (-\infty,\infty)$.

In the following, we compute the distribution of $y$ at time $t$ given the stationary state fluctuations of $k$ are described by Eq.~\eqref{k-eqn-21}. Since $y(t)$ is linear in $k(t)$ (Gaussian distributed noise), the distribution of $y(t)$ will also be Gaussian. Therefore, we only require its mean and variance to characterize its fluctuations. To this end, we solve Eq.~\eqref{y-eq}, yielding 
\begin{align}
 y(t) =y(0) -\sigma_0t - \sigma\int_0^t~dt'~k(t')\ ,
\end{align}
for the initial condition $y(0) = \ln x_0$. 
Averaging over 
trajectories of $k(t)$ given that its initial condition $k_0$ is drawn from the stationary distribution $p_{\rm ss}(k_0)$~\eqref{pk-ss}, we obtain the mean 
\begin{align}
 \mu_y\equiv  \langle {y(t)}\rangle  = 
 y(0) -\sigma_0t\ . \label{mu-y}
\end{align}
The variance of $y$ can be computed
via
\begin{align}
 \sigma_y^2\equiv \langle {\delta y(t)^2}\rangle  &= \sigma^2\int_0^t~dt_1\int_0^t~dt_2~\langle { k(t_1) k(t_2)}\rangle_{\rm ss}\nonumber \ ,\\
 &= -\sigma^2 H_1(t)\ . \label{sigma-y}
\end{align}
where we substituted the two-point correlation, $\langle { k(t_1) k(t_2)}\rangle_{\rm ss}$, from Eq.~\eqref{corr-k}. Further, $\delta y(t)\equiv y(t) - \langle {y(t)}\rangle$ is the deviation of $y$ from its mean value, and we defined
\begin{align}
    H_1(t)
    \equiv A t_p(\theta +2 t_p) (1 - t/t_p-e^{-t/t_p})\ . \label{H1-t}
\end{align}

\begin{figure}
    \centering
    \includegraphics[width=\textwidth]{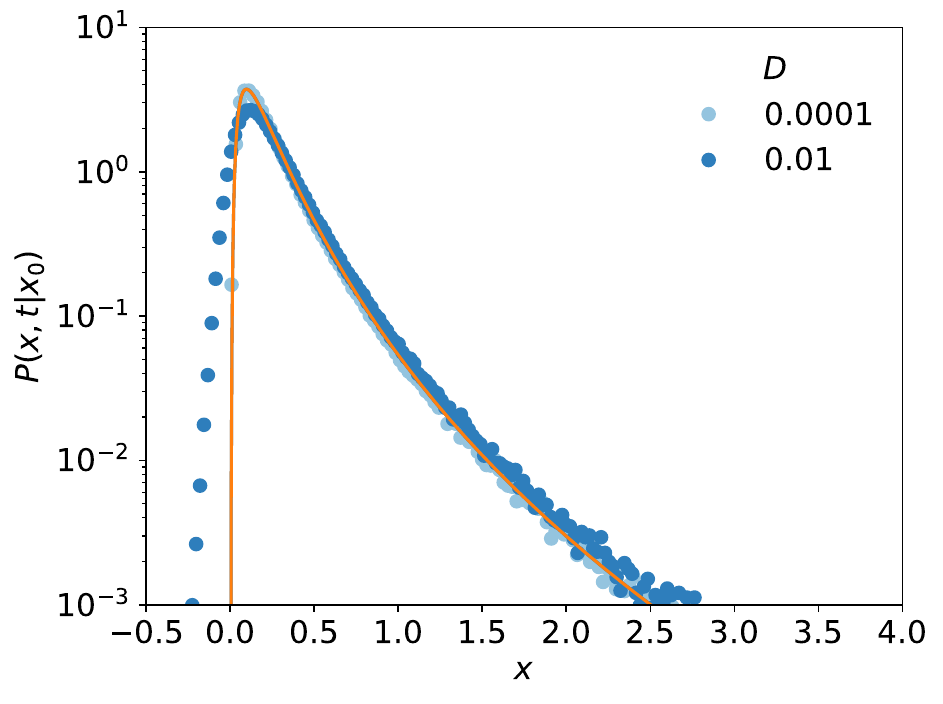}
    \caption{Comparison of the analytical probability density function for $D=0$~\eqref{ana-dis-deq0} and the numerically simulated distribution using Eq.~\eqref{eqns} for $D=10^{-4}, 10^{-2}$.  Other fixed parameters are $x_0 = 0.5$, $t_p=1.0$, $A=2$, $\theta =1.5$, $\sigma =0.5$, and $t=1.$  Solid line: Analytical result. Symbols: Numerical Langevin simulation (number of realizations: $10^6$).}
    \label{fig:comp-deq-and-d-neq}
\end{figure}

Given the $y$-distribution, $\pi(y,t|y(0))$, we obtain the position distribution while using the Jacobian of transformation of variables. 
It turns out that the positions' fluctuations are log-normal distributed, that is,
    \begin{align}
   P(x,t|x_0) =  \pi(y,t|y(0)) \bigg|\dfrac{dy}{dx}\bigg|=\dfrac{ \exp\bigg[-\dfrac{(\ln x-\mu_y)^2}{2\sigma_y^2}\bigg]}{x\sqrt{2\pi \sigma_y^2}}\ , \label{ana-dis-deq0}
\end{align}
for $x\geq 0$.  For $x\leq 0$, the above expression still holds by simply replacing $x\to |x|$. {\color{black}Clearly, the distribution does not approach a stationary state for any value of $\sigma_0$. The expression~\eqref{ana-dis-deq0} reveals that} 
the motion of the particle is restricted to either sides of the origin if the particle starts from $x_0\neq 0$. This behavior of the system can be intuitively understood as follows. In the absence of thermal noise [see Eq.~\eqref{eqns-21}], the position of the particle evolves as $\dot x(t) = -[\sigma_0 + \sigma k(t)]x(t)$. Therefore, 
a large positive value of $k(t)$ 
will bring the particle to origin, whereas 
a large negative 
value
pushes the particle to either $x=+\infty$ or $x=-\infty$, depending on whether the particle started from $x_0>0$ or $x_0<0$, respectively. Thus, the particle (for $D=0$)
is unable to cross the origin if it starts from $x_0\neq 0$.

Figure~\ref{fig:comp-deq-and-d-neq} shows the comparison of analytical expression 
for the probability density~\eqref{ana-dis-deq0} with numerical Langevin simulation performed for different thermal noise amplitude $D$. 
{\color{black}As expected, 
the departure of the numerical simulations' data for $D\neq 0$ from the analytical result for $D=0$ becomes evident as the value of $D$ increases, since for $D>0$ the system explores the full phase space, i.e., thermal fluctuations helps the particle to cross the origin.}

We further note that we can rewrite $P(x,t|x_0)$~\eqref{ana-dis-deq0} for $\sigma_0\neq 0$ as
\begin{align}
    P(z,t|z_0) =\dfrac{|\sigma_0| t}{\sqrt{2\pi \sigma_y^2}}\exp\bigg[-\dfrac{t^2\sigma_0^2(z- z_0 + 1)^2}{2\sigma_y^2}\bigg]\ ,\label{pz-t}
\end{align}
where we defined $z\equiv \frac{\ln |x|}{\sigma_0 t}$ and $z_0 \equiv \frac{\ln |x_0|}{\sigma_0 t}$. 
In the long time limit
$t\gg \sigma_0^{-1}\ln |x_0|$ (i.e., $z_0\ll 1$), 
the probability density function~\eqref{pz-t}  
assumes the large-deviation form~\cite{Hugo-1}
\begin{align}
    P(z,t)
    \approx e^{-t I(z)}\ 
\end{align}
with the large deviation function (LDF)
\begin{align}
    I(z) = \dfrac{\sigma_0^2(z+1)^2}{2 \sigma^2 A(\theta+2t_p)}\ .\label{ldf1}
\end{align}
Similarly, for $\sigma_0=0$, $P(x,t|x_0)$~\eqref{ana-dis-deq0} can rewritten as
\begin{align}
    P(z,t|z_0) =\dfrac{t}{\sqrt{2\pi \sigma_y^2}}\exp\bigg[-t^2\dfrac{(z-z_0)^2}{2\sigma_y^2}\bigg]\ ,\label{pz-t-2}
\end{align}
where we defined  we defined $z\equiv \frac{\ln |x|}{t}$ and $z_0 \equiv \frac{\ln |x_0|}{t}$. 
In the long time limit 
$t\gg \ln |x_0|$ (i.e., $z_0\ll 1$),   
we then obtain the large deviation form
\begin{align}
    P(z,t)
    \approx e^{-t I(z)}\ 
\end{align}
with the LDF
\begin{align}
    I(z) = \dfrac{z^2}{2 \sigma^2 A(\theta+2t_p)}\ .\label{ldf2}
\end{align}
The LDFs~\eqref{ldf1} and \eqref{ldf2}, respectively, describe the asymptotic form of the probability density functions (PDFs)~\eqref{pz-t} and \eqref{pz-t-2} in the long time limit: The LDF essentially describes that the deviation from the typical fluctuations (around the minimum of the LDF) are exponentially unlikely in the long-time limit~\cite{Hugo-1}.

\section{Presence of heat bath: $T\neq 0$}
\label{sec:Dneq0}
We now analyze the positions' fluctuations when $D\neq 0$. 
{\color{black}To proceed further, we rescale the position $x(t)$ in Eq.~\eqref{eqns-21} with the length scale $\sqrt{Dt_k}$ to write Eq.~\eqref{eqns-21} in the dimensionless form
\begin{subequations}
\label{eqns-2}
\begin{align}
    \dot x &= - [\sigma_0 + \sigma k(t)]x(t) + \sqrt{2}\eta(t) \label{x-eqn-2}\ , \\ 
        \dot k &= -\dfrac{k(t)}{t_p} + \sqrt{A\dfrac{\theta+2t_p}{t_p^2}}\xi(t) \label{k-eqn-2} \ .
\end{align}
\end{subequations}
In the following subsections, we} first investigate 
two particular limits of the persistence time $t_p$ entering in Eq.~\eqref{k-eqn-2}: These are, 1) $t_p\to 0$ (Gaussian white noise) and 2) $t_p\to \infty$ (quenched noise).
We then consider the case 
of general $t_p$. 
We here recall that the consideration of the different limits of $t_p$ is possible due to our choice of the prefactor in front of $\xi(t)$ [Eq.\eqref{k-eqn-2}].

\label{sec:Dneq0}
\subsection{White noise: $t_p\to 0$}
\label{sec:WN-main}
The limit $t_p\to 0$ leads to the white noise limiting behavior~[see first line of Eq.~\eqref{corr-k-lim}] of the full system, i.e., $k(t)$ is replaced by $\sqrt{A\theta} \xi(t)$ in Eq.~\eqref{x-eqn-2}. Given this, we herein compute the 
probability density function of this system.

The Fokker-Planck equation (following Stratonovich's rule)  corresponding to this system can be obtained as (see~\ref{sec:FP-WN} for its detailed derivation)
\begin{align}
    \dfrac{\partial P(x,t|x_0)}{\partial t} =  \dfrac{\partial}{\partial x} \bigg[(\sigma_0 -  a^2/2) x P(x,t|x_0)\bigg] +     \dfrac{\partial^2}{\partial x^2} \bigg[ (1 + a^2 x^2/2) P(x,t|x_0)\bigg]\ ,\label{FP-WN-main}
\end{align}
for the initial condition $P(x,t|x_0) = \delta(x-x_0)$ and $a \equiv  \sigma\sqrt{A\theta}$.
In the above equation~\eqref{FP-WN-main}, on the right-hand side, the terms proportional to $a^2$ are due to the stochastic white-noise stiffness. 
\begin{figure}
    \centering
    \includegraphics[width=\textwidth]{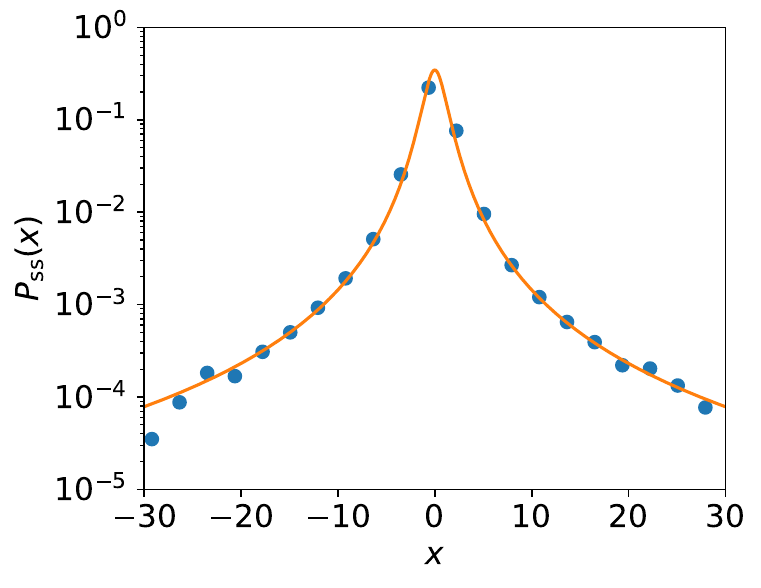}
    \caption{Comparison of analytical stationary state and numerical Langevin simulation data obtained at $t=50$ in the white noise limiting case ($t_p\to 0$). Other fixed parameters are $A = 1$, $\theta = 1.2$, $\sigma_0=1$, and $\sigma=1$. $t=50$. Solid line: Analytical result~\eqref{WN-ss-main}. Symbols: Numerical Langevin simulation.}
    \label{fig:WN-ss}
\end{figure}

The stationary state of the system can be obtained by setting the left hand side of Eq.~\eqref{FP-WN-main} equal to zero. It turns out that the stationary state probability density function is normalized only if $\sigma_0>0$, i.e., 
the system has a stationary state only if the underlying static harmonic potential is bounded (i.e., $\sigma_0>0$). This stationary distribution is given by
\begin{align}
    P_{\rm ss}(x) = \dfrac{a~2^{\frac{\sigma_0}{a^2}}}{\sqrt{\pi}}  \dfrac{\Gamma \left(\frac{1}{2}+\frac{\sigma_0}{a^2}\right)}{\Gamma \left(\frac{\sigma_0}{a^2}\right)} 
    \dfrac{1}{(2 + a^2 x^2)^{\frac{2\sigma_0 +a^2}{2a^2}}} \label{WN-ss-main}\ . 
\end{align} 
Thus, the positions' fluctuations in the stationary state have a power-law distribution and its tails decay as $|x|^{-(1+2\sigma_0/a^2)}$ as $|x|\to \infty$. Figure~\eqref{fig:WN-ss} shows the agreement of the analytical result~\eqref{WN-ss-main} with the numerical Langevin simulation data at $t=50$. Further, in the limit of vanishing stiffness' noise parameter (i.e., $a\to 0$), the stationary distribution~\eqref{WN-ss-main} becomes 
a Gibbs' distribution
\begin{align}
    P_{\rm ss}(x)\approx \mathcal{N}[1 - a^2 (\sigma_0/a^2 + 1/2) x^2/2] \to \mathcal{N} e^{- \sigma_0 x^2/2}\ ,
\end{align}
as expected, for the normalization constant $\mathcal{N}$.

Using this stationary state~\eqref{WN-ss-main}, we compute the $n$th position moment and find that  
\begin{align}
   \langle x^n\rangle_{\rm ss} = \frac{2^{\frac{n}{2}-1} \left[(-1)^n+1\right] a^{-n} \Gamma \left(\frac{n+1}{2}\right) \Gamma \left(\frac{\sigma_0}{a^2}-\frac{n}{2}\right)}{\sqrt{\pi } \Gamma \left(\frac{\sigma_0}{a^2}\right)}\ . \label{n-wn-moment}
\end{align}
Equation~\eqref{n-wn-moment} shows that the moments exist for $a^2 < 2\sigma_0/n$.

\subsection{Quenched noise: $t_p\to \infty$}
\label{sec:QN-main}
In the limit of long-persistence time ($t_p\to \infty$), the stochastic stiffness $k(t)$ is effectively constant in time  and behaves as a quenched Gaussian random variable [see the discussion below Eq.~\eqref{corr-k-lim}, and Eq.~\eqref{quench-dist} below] with zero mean and  $A$ variance [see Eqs.~\eqref{k-eqn-2},~\eqref{corr-k}]. [Notice that the quenched disorder appears in our model due to the prefactor (different compared to Ref.~\cite{Cocconi_2024}) in front of $\xi(t)$~\eqref{k-eqn}.] Therefore, for each given initial value of $k$ [drawn from the stationary distribution~$p_{\rm ss}(k)\propto e^{-k^2/(2A)}$], the probability density function of the particle's position is a Gaussian distribution
\begin{align}
    p(x,t;k|x_0) = \dfrac{1}{\sqrt{2 \pi \Sigma^2(t;k)} } \exp\bigg[-\dfrac{[x - \mu(t; k)]^2}{2 \Sigma^2(t;k)}\bigg]\ , \label{QN-fixed-k}
\end{align}
where the mean and variance, respectively, 
follows as 
\begin{align}
    \mu(t;k)& = x_0 e^{-(\sigma_0 + \sigma k)t}\ ,\\
    \Sigma^2(t;k) &=  \dfrac{1 - e^{-2  (\sigma_0 + \sigma k) t}}{\sigma_0 + \sigma k} \ .
\end{align}
\begin{figure}
    \centering
    \includegraphics[width=\textwidth]{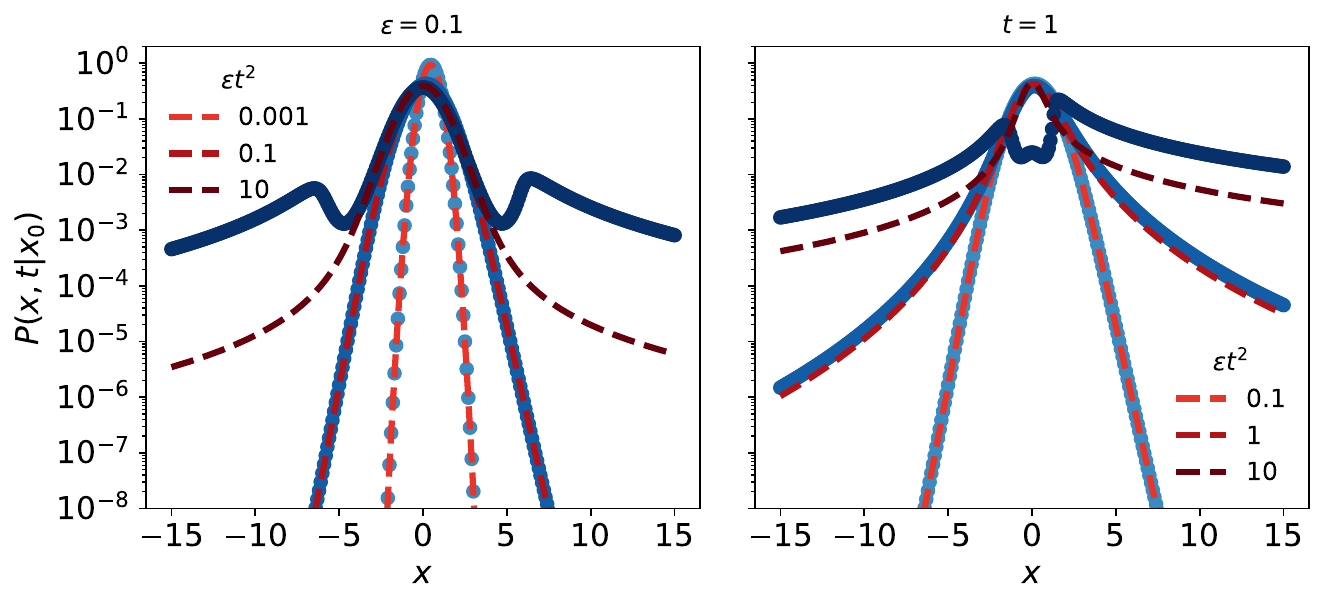}
    \caption{Comparison of saddle-point (blue symbol) approximated distribution~\eqref{p-x-sd-pt} with the exact (red dashed line) distribution~\eqref{before-sd-pt} in the quenched noise case ($t_p\to \infty$). We fixed $\sigma_0=1$ and the initial position $x_0 =0.5$. Color intensity increases with increasing $\epsilon t^2$.}
    \label{fig:sp-ex-comp}
\end{figure}

We are now in the position to obtain the distribution of $x$ at time $t$ 
by averaging $p(x,t;k|x_0)$ over the distribution of $k$-values, yielding
\begin{align}
    P(x,t|x_0) &= \int_{-\infty}^{+\infty}~dk~p_{\rm ss}(k)~p(x,t;k|x_0)\label{quench-dist}\\
   &= \frac{1}{2\pi \sigma\sqrt{A}t}\int_{0}^{+\infty}~ \frac{dy}{y}\sqrt{\frac{\ln y}{t(y^2-1)}}e^{-\frac{(\ln y + \sigma_0t)^2}{2A\sigma^2 t^2}} e^{-\frac{\ln y}{t}\frac{(x - x_0 y)^2}{2(y^2-1)}} \label{px-eq1}\ .
\end{align}
To reach Eq.~\eqref{px-eq1} from 
Eq.~\eqref{quench-dist}, we made a change of variables such that $k = -(\sigma_0t + \ln y)/(\sigma t)$. 
Notice that, unlike the case of a harmonic oscillator with a fixed (non-random) stiffness--where the distribution remains Gaussian at all times--here, due to the quenched stochastic stiffness, the resulting distribution is not necessarily Gaussian (see dashed lines in Fig.~\ref{fig:sp-ex-comp}).

It is readily possible to evaluate the 
integral in Eq.~\eqref{px-eq1} numerically. Still, it is interesting to
find 
an approximation  
of the above integral~\eqref{px-eq1} when the external noise strength, $\sigma^2A$, is vanishingly small. This can done by using the saddle-point method. To this end, we rewrite the above integral~\eqref{px-eq1} as
\begin{align}
    P(x,t|x_0) = \frac{1}{2\pi \sqrt{\epsilon} t} \int_0^\infty~dy~\exp\bigg[-\dfrac{1}{2\epsilon t^2}f(y;x,t,x_0,\epsilon)\bigg] \label{before-sd-pt}
\end{align}
where we defined $\epsilon \equiv \sigma^2 A$, and 
\begin{align}
  f(y;x,t,x_0,\epsilon) 
   &\equiv \epsilon t^2\ln \left(\frac{ ty^2\left(y^2-1\right)}{\ln y}\right) +\epsilon t \ln y  \frac{(x-x_0 y)^2}{\left(y^2-1\right)}+(\sigma_0t+\ln y)^2\ .
\end{align}
Considering $\epsilon t^2$ as a small parameter (i.e., $\epsilon t^2 \ll 1$), we can approximate the above integral using the saddle-point method. 
This gives 
\begin{align}
    P(x,t|x_0)&\approx \dfrac{\exp \left(-\frac{f(y^*)}{2 t^2 \epsilon }\right)}{\sqrt{4\pi f''(y^*)}}\left[1 + \text{Erf}\left(y^*(x) \sqrt{\frac{f''(y^*)}{4 \epsilon t^2}}\right)\right]\ , \label{p-x-sd-pt}
\end{align}
where $y^*\equiv y^*(x,t,x_0,\epsilon)$ is the saddle-point obtained by solving 
\begin{align}
    \partial_y f(y;x,t,x_0,\epsilon)|_{y=y^*} =0\ .
\end{align}
For convenience, in Eq.~\eqref{p-x-sd-pt} we 
used the notation $f(y^*)\equiv f(y^*;x,t,x_0,\epsilon)$ and $f''(y) \equiv \partial^2_y f(y;x,t,x_0,\epsilon)|_{y=y^*}$.  

Figure~\ref{fig:sp-ex-comp} shows the comparison of the saddle-point result~\eqref{p-x-sd-pt} with the exact distribution~\eqref{before-sd-pt} for different values of $\epsilon t^2$ 
and fixed $\sigma_0=1$ (similar findings occur for $\sigma_0\leq 0$). 
The data shows that the saddle-point approximation becomes increasingly better as $\epsilon t^2$ decreases.
Finally, we here remark that for $\sigma_0>0$ and in the long-time limit, the system will approach a stationary state in the limit $\epsilon\to 0$. 
For other cases of $\sigma_0$, the system does not have a stationary state even in this limit ($\epsilon\to 0$) due to the non-positive value 
of stiffness $k$, as expected.

\subsection{General $t_p$}
\label{sec:gen_tp}
In this section, we treat the full model~\eqref{eqns-2} with $D\neq 0$. 
To compute the 
the probability density function of particle's position, 
we solve the Fokker-Planck equation for each realization of $k(t)$, given as 
\begin{align}
    \dfrac{\partial p(x,t;k(t))}{\partial t}  = \dfrac{\partial }{\partial x} \big[(\sigma_0 + \sigma k(t)]x p(x,t;k(t))\big] +\dfrac{\partial^2 p(x,t;k(t))}{\partial x^2} \ , \label{FPE}
\end{align}
for initial condition $p(x,0;k_0) = \delta(x-x_0)$ for all $k_0\equiv k(0)$. Notice here that $k_0$ are distributed according to the stationary distribution $p_{\rm ss}(k_0)$~\eqref{pk-ss}. Therefore, once $p(x,t;k(t))$ is known, we 
perform the average 
\begin{align}
    P(x,t|x_0) = \langle p(x,t;k(t))\rangle_{\{k(t)\}} \ ,\label{ann-avg}
\end{align}
over the  
ensemble of trajectories of $k(t)$ emanating from $k_0$, where $k_0$ is drawn from $p_{\rm ss}(k_0)$~\eqref{pk-ss}

To solve the Fokker-Planck equation~\eqref{FPE} for a given realization of $k(t)$, we use the method of subordination~\cite{MOS-1,MOS-2}. 
The implies the substitution 
\begin{align}
    p(x,t;k(t)) =  f(t)~Q(z(x,t), \tau(t))\ , \label{p-factor}
\end{align}
where 
\begin{subequations}
\label{fzt}
\begin{align}
    f(t) &\equiv  e^{\sigma_0t+\sigma\int_0^t~ds~k(s)}\ ,\label{fteqn}\\
    z(x,t) &\equiv x  f(t)\ \label{eqn-zt},\\
    \tau(t) &\equiv \int_0^t~ds~ f^2(s) \label{tau} \ ,
\end{align}
\end{subequations}
for the initial condition $z(x,0)=x_0$, in Eq.~\eqref{FPE}. 
With this substitution, the Fokker-Planck equation~\eqref{FPE} translates into a diffusion equation
\begin{align}
    \partial_\tau Q(z,\tau) =\partial_z^2 Q(z,\tau)\ , \label{w-diff}
\end{align}
with the stochastic time $\tau$~\eqref{tau}. 
Equation~\eqref{w-diff} is supplemented 
by the initial condition $Q(z,0) = \delta(z - x_0)$. For convenience,  we henceforth drop the explicit dependence of $x$ and $t$ from $z$ and $\tau$.

The solution of Eq.~\eqref{w-diff} can be obtained using the Fourier-Transform
\begin{align}
    Q(z,\tau)
    =\dfrac{1}{2\pi}\int_{-\infty}^{+\infty}~dp~e^{- p^2 \tau} e^{ip(z-x_0)}=\dfrac{1}{2\pi}\int_{-\infty}^{+\infty}~dp~e^{-ipx_0}e^{ipxf(t)}~ e^{- p^2\tau(t)} \ , \label{wsol}
\end{align}
where $p$ is the conjugate variable with respect to $z$, and we have substituted $z$ from Eq.~\eqref{eqn-zt}.  

Substituting ~\eqref{wsol} for $Q(z,\tau)$ into Eq.~\eqref{p-factor}, and 
writing the annealed average~\eqref{ann-avg} over the ensemble of trajectories of $k(t)$, we have
\begin{align}
       P(x,t|x_0)=\bigg\langle \int_{-\infty}^{+\infty}~~dp~f(t)\dfrac{e^{-ipx_0}}{2\pi}~ e^{- p^2 \tau(t)} e^{ipxf(t)}\bigg\rangle_{\{k(t)\}} \ . \label{P-full-wo-avg}
\end{align}
Henceforth, for notational simplicity, we drop the subscript ${\{k(t)\}}$ from the angled brackets. Due to the complicated structure of $f(t)$~\eqref{fteqn} as well as $\tau(t)$~\eqref{tau} entering in the exponents in Eq.~\eqref{P-full-wo-avg}, the 
computation of the ensemble average~\eqref{P-full-wo-avg} to obtain the probability density function is not straightforward. Nonetheless, Eq.~\eqref{P-full-wo-avg} 
provides the basis to compute the $n$th moment of the position of the particle. To this end, we multiply $x^n$ on both sides of Eq.~\eqref{P-full-wo-avg} and integrate from $x=-\infty$ to $x=+\infty$. 
This gives
\begin{align}
    \langle x^n \rangle & = \sum _{q=0, q\in \mathbb{Z}^+}^{n/2} \binom{n}{2q}\frac{4^q}{\sqrt{\pi}}~\Gamma\left(\frac{1+2q}{2}\right)  x_0^{n-2q}\langle f^{-n} \tau^{q}  \rangle \ , \label{nth-mom}
\end{align}
where, for convenience, we have dropped the explicit dependence of $t$ from $f(t)$ and $\tau(t)$. Here, we remark that our results in this section generalizes that of Ref.~\cite{Cocconi_2024}, by considering the non-zero initial position ($x_0\neq 0$). Moreover, the method used in this paper (method of substitution) is different than that of used in Ref.~\cite{Cocconi_2024}.

For $n=0$, 
both sides give unity, as expected. 
To compute the $n$th moment, we have to 
evaluate the right-most term, i.e., the average. In the following, we discuss the method to compute this quantity.

We start from the explicit expression for the average, 
\begin{align}
    \langle f^{-n} \tau^{q} \rangle &= \bigg\langle e^{-n \sigma_0 t} e^{-n\sigma \int_0^t~ds~k(s)}\prod_{j = 1}^q~\int_0^t~ds_j~e^{2 \sigma_0s_j} e^{2\sigma \int_0^{s_j}~da_j~k(a_j)}  \bigg\rangle \nonumber\\
      & =  e^{-n \sigma_0 t} \bigg(\prod_{j = 1}^q~\int_0^t~ds_j\bigg)~e^{2 \sigma_0 \sum_{\ell = 1}^q s_\ell}\bigg\langle e^{-\sigma \big[n \int_0^t~ds~k(s) - 2 \sum_{m=1}^q\int_0^{s_m}~da_m~k(a_m)\big]}  \bigg\rangle \ .\label{ftau-first}
\end{align}

We now use the fact that for a random variable $Y$, the cumulant generating function is $\ln \langle e^{-\sigma Y} \rangle = \sum_{m=1}^\infty~\frac{(-\sigma)^m}{m!}\llangle Y^m \rrangle$, where the double angled brackets indicates the cumulant. Thus,
\begin{align}
    \langle f^{-n} \tau^{q} \rangle & =  e^{-n \sigma_0 t} \bigg(\prod_{j = 1}^q~\int_0^t~ds_j\bigg)~e^{2 \sigma_0 \sum_{\ell = 1}^q s_\ell}~\times\nonumber\\
    &\times \exp\bigg[{\sum_{m=1}^\infty~\frac{(-\sigma)^m}{m!}~\bigg\llangle \bigg[n \int_0^t~ds~k(s) - 2 \sum_{m=1}^q\int_0^{s_m}~da_m~k(a_m)\bigg]^m\bigg\rrangle}\bigg]\ . \label{ftau-exct}
\end{align}
We emphasize that the above expression is exact and does not depend on the underlying dynamics of $k(t)$.

Since the term {\color{black}inside} the square bracket is Gaussian random variable [as in our case~\eqref{k-eqn-2}],  
the cumulants of order larger
than 2 (i.e., $m>2$) 
are zero. Further, the term corresponding to $m=1$ is also zero as the first cumulant or the mean of $k(t)$ is zero. 
Equation~\eqref{ftau-exct} thus reduces to
\begin{align}
    \langle f^{-n} \tau^{q} \rangle & =  e^{-n \sigma_0 t} \bigg(\prod_{j = 1}^q~\int_0^t~ds_j\bigg)~e^{2 \sigma_0 \sum_{\ell = 1}^q s_\ell}~\times \nonumber\\
    &\times \exp\bigg[{~\frac{\sigma^2}{2}~\bigg\llangle \bigg[n \int_0^t~ds~k(s) - 2 \sum_{m=1}^q\int_0^{s_m}~da_m~k(a_m)\bigg]^2\bigg\rrangle}\bigg]\ . \label{ftau-leading}
\end{align}
We here remark that in some cases, the dynamics of $k(t)$ 
may be governed by a complex stochastic process that reaches a stationary state at long times. In such situations, calculating fluctuations in the position (position's moments) is not straightforward. However, when the stochastic stiffness is weak (i.e., $\sigma \to 0$), we can still compute the moments of the position by using the average $\langle f^{-n} \tau^{q} \rangle$ from Eq.~\eqref{ftau-leading} in Eq.~\eqref{nth-mom}. This average captures the leading contributions of order $\sigma^2$, assuming that higher-order cumulants $\llangle k^n(t) \rrangle$ remain finite for $n > 2$. Additionally, if the stationary process $k(t)$ has a nonzero mean $\langle k(t) \rangle_{\rm ss}$, we can shift it to zero by redefining $\sigma_0$ as $\sigma_0 + \sigma \langle k(t) \rangle_{\rm ss}$.

Coming back to Eq.~\eqref{ftau-leading}, inside the exponential, we expand the square bracket and use the expression~\eqref{corr-k} for $\llangle k(t)k(s)  \rrangle$ since $k(t)=0$.
Performing the integrals, we obtain
 \begin{align}
   \langle f^{-n} \tau^{q} \rangle   &=  e^{-n \sigma_0 t} \bigg(\prod_{j = 1}^q~\int_0^t~ds_j\bigg)~e^{2\sigma_0 \sum_{\ell = 1}^q s_\ell}e^{-\frac{\sigma^2}{2}[n^2 H_1(t)+4 \sum_{m_1=1}^q\sum_{m_2=1}^q H_2(s_{m_1},s_{m_2})  -4n \sum_{m=1}^q H_2(t,s_{m})]}\ ,  \label{ftau-fin}
\end{align}
where we defined
\begin{align}
  H_2(s_{1},s_{2})\equiv \frac{1}{2} A t_p (\theta +2 t_p) \left[1-2\frac{\min (s_{1},s_{2})}{t_p}+e^{-\frac{|s_{1}-s_{2}|}{t_p}}-e^{-\frac{s_{1}}{t_p}}-e^{-\frac{s_{2}}{t_p}}\right]\ ,
\end{align}   
and $H_1(t) = H_2(t,t)$ [$H_1(t)$ is given in Eq.~\eqref{H1-t}]. Finally, substituting Eq.~\eqref{ftau-fin}  
into the right-hand side of Eq.~\eqref{nth-mom}, we obtain the expression for
all 
moments.

In the following, we obtain  the first four positional moments using Eq.~\eqref{nth-mom}. For $n=1$, we get
\begin{align}
       \langle x \rangle&=x_0\langle f^{-1}(t)\rangle= x_0 e^{-\sigma_0 t} e^{-\sigma^2 H_1(t)/2}\ .\label{1stmom}
\end{align}
In the long-time limit, $H_1(t)$ behaves as $H_1(t) \approx -A (\theta + 2t_p) t$.
Therefore, the first moment reach a stationary value $\langle x \rangle_{t\to\infty} =0$,
when the parameter combination is such that $A \sigma^2 (\theta + 2t_p)\leq 2\sigma_0$. Thus, the mean position always diverges for the unbounded system $\sigma_0 \leq 0$ for non-zero initial condition $x_0\neq 0$.

Let us now compute the second moment ($n=2$)  using Eq.~\eqref{nth-mom}. 
We find
\begin{align}
       \langle x^2 \rangle &=x_0^2 \langle f^{-2}(t)\rangle + 2 \langle f^{-2}(t)\tau(t)\rangle=x^2_0 e^{-2\sigma_0t}e^{-2\sigma^2 H_1(t)} +2 \underbrace{\int_0^t~ds~ e^{-2\sigma_0s}~e^{-2\sigma^2 H_1(s)}}_{\mathcal{I}_1}\label{def-I1}\\
       &=x^2_0 e^{-2\sigma_0t}e^{-2\sigma^2 H_1(t)} + 2t_p  e^{-\alpha }(-\alpha )^{-b} \left[\Gamma \left(b,-e^{-\frac{t}{t_p}} \alpha \right)-\Gamma (b,-\alpha )\right]\ , ~{\rm for}~t_p>0 \label{main:x2-t-val}
\end{align}
where $\Gamma(b,y) = \int_y^\infty~x^{b-1}e^{-x}$ is the incomplete gamma function, and we defined $\alpha \equiv 2 \sigma^2 A t_p(\theta + 2 t_p)$ and $b \equiv 2\sigma_0t_p - \alpha$ (see~\ref{sec:I1} for the computation of $\mathcal{I}_1$). In the long-time limit, the second moment reaches a stationary value 
\begin{align}
   \langle x^2 \rangle_{t\to\infty} = 2t_p  e^{-\alpha }(-\alpha )^{-b} \left[\Gamma \left(b\right)-\Gamma (b,-\alpha )\right]\ , \label{x2-ltime} 
\end{align}
when $b>0$. Therefore, the second moment reaches a stationary value if $A\sigma^2(\theta + 2 t_p) < \sigma_0$.

\begin{figure}
    \centering
    \includegraphics[width=\textwidth]{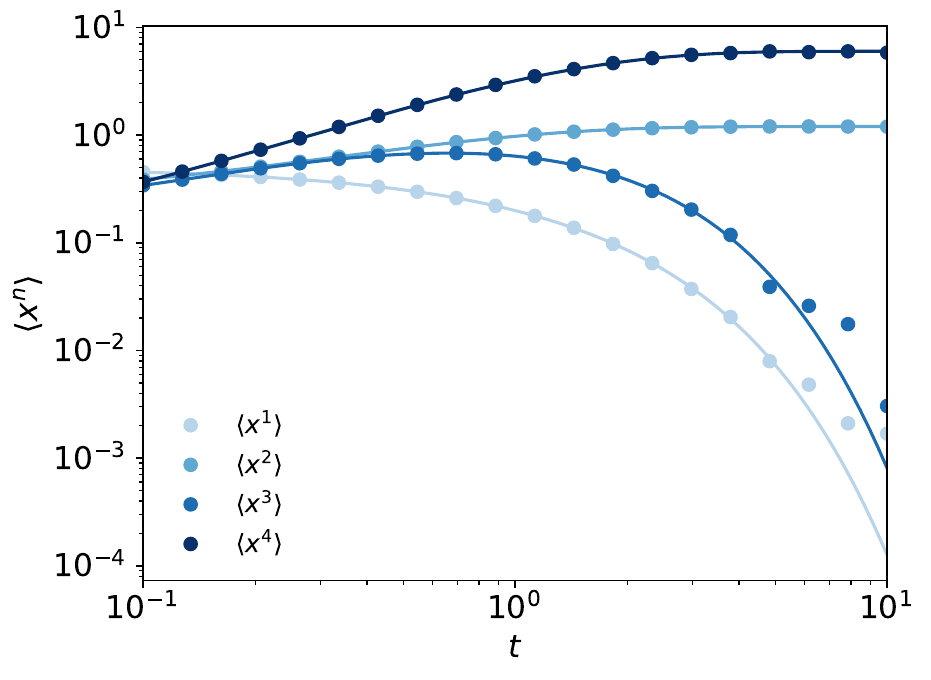}
    \caption{Comparison of analytical results for the moments with the numerical simulations. We take the parameters $x_0=0.5, \sigma =0.65,~t_p=0.8,~\theta =0.6,~A=0.4$ such that $A \sigma ^2 (\theta +2 t_p) = 0.3718$. $dt = 10^{-4}$. Number of realizations $10^6$. Solid lines: Analytical results~\eqref{1stmom}, \eqref{main:x2-t-val}, \eqref{3thmom}, and \eqref{4thmom}. Circles: Numerical Langevin simulations. }
    \label{fig:pos-mom}
\end{figure}

The third moment ($n=3$) 
follows as
\begin{align}
   \langle x^3 \rangle
 &= x_0^{3}\langle f^{-3} \rangle + 6 x_0~\langle f^{-3} \tau^{1}  \rangle\nonumber\\
 &=x_0^{3} e^{-3\sigma_0 t} e^{-9\sigma^2/2 H_1(t)} +6x_0~e^{-3\sigma_0 t}e^{-9\sigma^2/2 H_1(t)} \underbrace{\int_0^t~ds e^{2\sigma_0 s} e^{-2\sigma^2[H_2(s,s)  -3H_2(t,s)]}}_{\mathcal{I}_2}\label{def-I2}\\
 &=x_0^{3} e^{-3\sigma_0 t} e^{-9\sigma^2/2 H_1(t)} +6x_0~t_p~e^{\alpha/2} e^{-\frac{3 \alpha}{2}  e^{-t/t_p}} e^{-3\sigma_0 t}e^{-9\sigma^2/2 H_1(t)}~\times\nonumber\\
 &\times  \sum_{m=0}^\infty \dfrac{(3\alpha/2)^m e^{-mt/t_p} }{m!} \left[e^{\frac{t (b-\alpha +m)}{t_p}} E_{b+m-\alpha +1}\left(\frac{\alpha e^{-\frac{t}{t_p}}}{2} \right)-E_{b+m-\alpha +1}\left(\frac{\alpha }{2}\right)\right],\label{3thmom}
 \end{align}
 where $\alpha$ and $b$ are defined below Eq.~\eqref{main:x2-t-val}, and $E_n(y)=\int_1^{\infty }~dx~x^{-n} e^{-xy}$ is the exponential integral function. Here, $\mathcal{I}_2$ is evaluated in \ref{sec:I2}. 
In the long-time limit, both first and second terms go to zero if $A\sigma^2(\theta + 2 t_p) < 2 \sigma_0/3$; therefore, $\langle x^3 \rangle_{t\to\infty} =0$.

Finally, we compute the fourth moment ($n=4$)
\begin{align}
\langle x^4 \rangle 
&= x_0^{4}\langle f^{-4}\tau^0 \rangle + 12  ~x_0^2~\langle f^{-4} \tau^{1}  \rangle + 12\langle f^{-4} \tau^{2}  \rangle\nonumber\\
&=x_0^{4} e^{-4\sigma_0 t}e^{-8\sigma^2 H_1(t)} + 12 x_0^2 e^{-4\sigma_0 t}e^{-8\sigma^2 H_1(t)}\int_0^t~ds~e^{2\sigma_0s} e^{-2\sigma^2[H_2(s,s) - 4 H_2(t,s)]}~+\nonumber\\
&+12 e^{-4\sigma_0 t}e^{-8\sigma^2 H_1(t)} \int_0^t~ds_1\int_0^t~ds_2~e^{2\sigma_0(s_1+s_2)}e^{-\sigma^2/2[4 \sum_{m_1=1}^2\sum_{m_2=1}^2H_2(s_{m_1},s_{m_2}) - 16 \sum_{m=1}^2H_2(t,s_m)]}\ .\label{4thmom}
\end{align}
While it is not straightforward to analyze the convergence criteria for the third integral, from the first two terms we find that these terms 
saturate at long times if $A\sigma^2(\theta + 2 t_p)<\sigma_0/2$.
We numerically checked that the fourth moment converges with this same condition. Thus, analyzing the first four moments, we find that these moments exist when $A\sigma^2(\theta + 2 t_p)<2\sigma_0/n$, where $n$ is the order of the moment. 
We hypothesize that this 
is true for any $n$, as in the case of white noise limit $t_p\to 0$ [see Eq.~\eqref{n-wn-moment}].
To verify 
expressions~\eqref{1stmom}, \eqref{main:x2-t-val}, \eqref{3thmom}, and \eqref{4thmom}, we have performed numerical Langevin simulations.  Figure~\ref{fig:pos-mom} shows a good agreement of the analytical results 
for each moment with the numerical Langevin simulations.

{\color{black}Finally, Table~\ref{table-1} summarizes the conditions under which the stationary state exists for different cases discussed above.}
\begin{table}[h!]
\centering
\begin{tabular}{|c|c|}
\hline
\textbf{Thermal} & \textbf{Stationary state} \\
 \textbf{fluctuations}&\\
\hline 
&\\
         $D = 0$ & No stationary state exists for any value of the static trap's parameter $\sigma_0$~\eqref{ana-dis-deq0}.\\
         &\\
         \hline
         &\\
        & $t_p\to 0$: 1) The stationary state exists for $\sigma_0>0$~\eqref{WN-ss-main}.~~~~~~~~~~~~~~~~~~~~~~~~~~~~~\\
         &2) The stationary $n$th moment exists for~$A\sigma^2\theta<\sigma_0/n$~\eqref{n-wn-moment}.\\
         &\\
         $D\neq 0$&$t_p\to \infty$: In the limit $\epsilon\to 0$, the stationary state exists for  $\sigma_0>0$ (Sec.~\ref{sec:QN-main}).\\
         &\\
         &General $t_p$: The stationary $n$th moment exists for $A\sigma^2(\theta+2t_p)<\sigma_0/n$.~~~~~~~\\
         &\\
\hline
\end{tabular}
\caption{{\color{black}Existence of a stationary state and the $n$th moment are discussed for different cases.}}\label{table-1}
\end{table}

\section{Thermodynamics}
\label{sec:thermo}
In this work, we discuss a system where the stiffness of the harmonic trap is a fluctuating quantity.  
As a consequence of these temporal stochastic fluctuations, the particle is driven away from equilibrium.  
We now analyze thermodynamic quantities, such as work performed on the system and heat exchanged by the system with the heat bath, as functions of time.

We start by considering the internal energy of the particle
$U(x,t;k)$, where $k$ is the time-dependent control parameter. Taking a time-derivative of 
the internal energy gives the rate of net change of internal energy.
This occurs due to rate of change of the particle's position and due to rate of change of the control parameter on a single stochastic trajectory level~\cite{sekimoto2010stochastic}: 
\begin{align}
    \dfrac{dU(x,t;k)}{dt} = \underbrace{\dfrac{\partial U(x,t;k)}{\partial x} \dot x}_{\dot{q}} +  \underbrace{\dfrac{\partial U(x,t;k)}{\partial k} \dot k}_{\dot{w}}\ , \label{netru}
\end{align}
where the first quantity on the right-hand side is the rate of heat entering in the system from the 
bath, while the second quantity is the rate of work performed on the system due to change 
of the control parameter $k$. (Notice that $q>0$ and $w>0$, respectively, correspond to the situation when  
heat flows from the bath to the system, and  
work is performed on the particle.) For our 
system described in Eq.~\eqref{eqns-2}, the system's internal energy in the reduced unit is given by
\begin{align}
    U(x,t;k) = \dfrac{1}{2}[\sigma_0 + \sigma k(t)]x^2\ .
\end{align}

Therefore, the work along a single stochastic trajectory 
follows from Eq.~\eqref{netru} as 
\begin{align}
    w(t) = \sigma\int_0^t~ds~\dfrac{\dot k}{2} x^2\ . \label{w-def}
\end{align}
Notice that $w$ goes to zero as $\sigma \to 0$, as it should be, since 
$\sigma\to 0$ corresponds to an equilibrium situation. 
Furthermore, the heat taken by the particle from the bath follows from Eq.~\eqref{netru} as
\begin{align}
    q(t) = \int_0^t~ds~[\sigma_0 + \sigma k(t)] x \dot{x}\ . \label{ht-1]}
\end{align}
Henceforth, for convenience, we drop the explicit mention of the time-dependence of $U(x,t;k)$, $w(t)$ and $q(t)$, and simply write as $U(x;k)$, $w$ and $q$.

Both work $w$ and heat $q$ are fluctuating quantities, 
that is, their values vary from one trajectory $[\{x, k\}]_{0<t'\leq t}$ to another. Therefore, it 
would be interesting to compute the full fluctuations [probabilities density function (PDF)] of these quantities. However, the computation of PDFs 
is not straightforward. Nevertheless, we can exactly compute the average of these quantities for all time. (Notice that Ref.~\cite{Cocconi_2024} discusses the entropy production only in the stationary state.)

In the following, we first show the calculation 
for the average work from Eq.~\eqref{w-def}. To this end, we perform 
an integration by parts
\begin{align}
    w = \dfrac{\sigma}{2}\bigg[k(s) V(s)\bigg]_0^t - \sigma\int_0^t~ds~\dfrac{1}{2}k(s)\dot{V}\ ,
\end{align}
where we defined $V(t) \equiv x^2(t)$.
The average work is obtained by performing the ensemble average over trajectories of $x$ and $k$, i.e., $\mathcal{W}\equiv  \langle w\rangle_{\{x,k\}}$.

We first perform the average over $x$ for a given realization of $k(t)$. 
Assuming that the particle starts from a fixed location $x_0$, we have
\begin{align}
    W = \dfrac{\sigma}{2}\bigg[k(t) \mathcal{V}(t) - k_0 V(0)\bigg] - \sigma\int_0^t~ds~k(s)\dfrac{\dot{\mathcal{V}} }{2}\ ,\label{w-def-2}
\end{align}
where we defined $W\equiv \langle w\rangle_{\{x\}}$ and  $\mathcal{V}(t)\equiv \langle V(t)\rangle_{\{x\}}$, where $\mathcal{V}$ is the second moment of the particle's position for a given realization of $k(t)$ starting from $k_0$. Notice that the initial condition $k_0$ is distributed according to $p_{\rm ss}(k_0)$~\eqref{pk-ss}.

We proceed by multiplying $x^2$ on both sides of Fokker-Planck equation~\eqref{FPE} and integrating from $x=-\infty$ to $x=+\infty$.  
This gives the time evolution of the second positional moment $\mathcal{V}$ for each realization of $k(t)$:
\begin{align}
    \dot{\mathcal{V}} (t) = -2[\sigma_0 + \sigma k(t)] {\mathcal{V} (t)} + 2\ . \label{var-eqn}
\end{align}

Substituting $\dot{\mathcal{V}} (t)$~\eqref{var-eqn} on the right-hand side of Eq.~\eqref{w-def-2}, and performing the average over noise realization of $k(t)$, we 
obtain the average work
\begin{align}
    \mathcal{W}= \dfrac{\sigma}{2}\langle k(t) \mathcal{V}(t) \rangle_{\{k\}} 
+ \sigma \sigma_0 \int_0^t~ds~\langle k(s) {\mathcal{V} (s)}\rangle_{\{k\}}  +\sigma^2 \int_0^t~ds~\langle k^2(s) {\mathcal{V} (s)}\rangle_{\{k\}}  \ , \label{lbo}
\end{align}
where we substituted the stationary state $\langle k_0\rangle = 0$. To compute the right-hand side of~\eqref{lbo}, we compute $\mathcal{V} (t)$  by solving Eq.~\eqref{var-eqn}. This yields
\begin{align}
    \mathcal{V} (t) =    V(0) e^{-2 [\sigma_0 t + \sigma \int_0^t~ds~k(s)]} + 2\int_0^t~dt'e^{-2 [\sigma_0 (t-t') + \sigma \int_{t'}^t~ds~k(s)]}\ , \label{v-eqn-1}
\end{align}
where $V(0)=\mathcal{V} (0)$ since the particle starts from a fixed location $x=x_0$.

For simplicity, we 
now assume that the situation where the particle starts from the origin at time $t=0$, i.e., $x_0=0$. Then, Eq.~\eqref{v-eqn-1} becomes
\begin{align}
    \mathcal{V} (t) =    2\int_0^t~dt'e^{-2[\sigma_0 (t-t') + \sigma \int_{t'}^t~ds~k(s)]}\ . \label{sec-mom-v-eqn}
\end{align}
With this, we can compute the averages on the right-hand side of Eq.~\eqref{lbo}. The explicit calculations of each of the terms  
is cumbersome. Therefore, we relegate the computational details in the \ref{app:work} and present here the final expressions as 
\begin{align}
   \langle k(s) {\mathcal{V} (s)}\rangle_{\{k\}} &=-4\sigma \int_0^s~du~~H_3(u)~e^{-2\sigma_0 u}e^{-2\sigma^2 H_1(u)}\ , \label{main:ku}\\
\langle k^2(s) {\mathcal{V} (s)}\rangle_{\{k\}} 
&=2 \int_0^s~du~\bigg[A \dfrac{(\theta+2 t_p)}{2t_p} +4 \sigma^2 H_3^2(u) \bigg] e^{-2\sigma_0 u}e^{-2\sigma^2 H_1(u)} \ , \label{main:k2u}
\end{align}

where we defined 
\begin{align}
    H_3(u) \equiv \frac{1}{2} A (\theta +2 t_p) \left(1-e^{-\frac{u}{t_p}}\right)\ .
\end{align}
Substituting the results from Eqs.~\eqref{main:ku} and \eqref{main:k2u} into Eq.~\eqref{lbo}, we obtain the expression for the average work in the integral form. These integrals can be exactly calculated
(see~\ref{app:w0w1w2}). We here report the long-time result. 
\begin{figure}
    \centering
    \includegraphics[width=\textwidth]{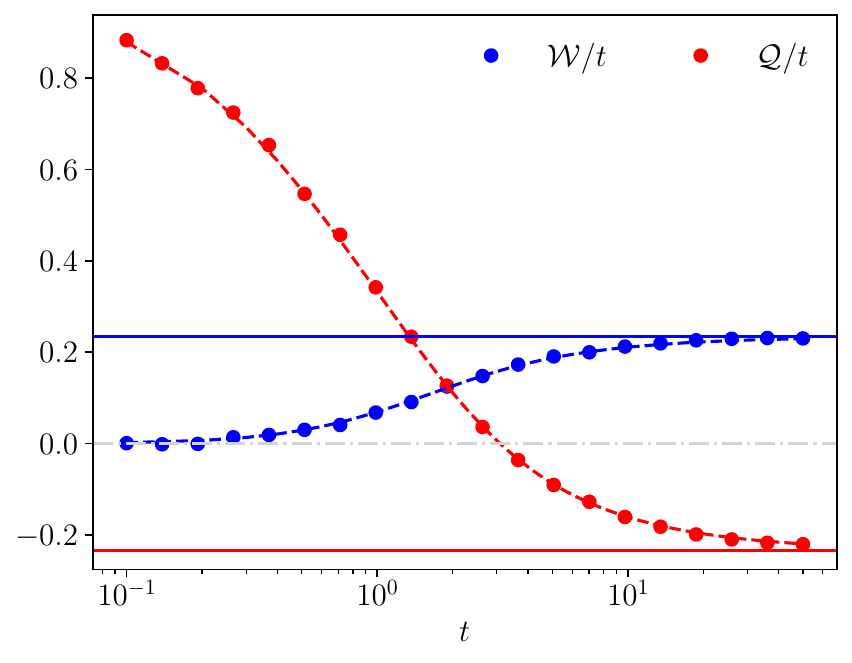}
    \caption{Average work per time (blue) and average heat per time (red), each as a function of observation time. Symbols: Numerical Langevin simulation. Dashed line: Analytical results~\eqref{lbo}~and~\eqref{ana-heat}. Solid horizontal line: Analytical result in the long-time limit~\eqref{scwk-lt} (blue) and the negative of the long-time limit expression~\eqref{scwk-lt} (red). The parameter are $A = 1.8$, $\theta = 2.5$, $t_p=0.5$, $\sigma_0=1$, $\sigma= 0.23$, and $dt=10^{-4}$. Number of realizations: $10^4$. }
    \label{fig:avg-work}
\end{figure}
In this long-time limit (for $b>0$), the scaled average work approaches 
[see Eqs.\eqref{si:lbo}, \eqref{w0bt_asyp}, \eqref{w1bt-syp}, and \eqref{w2bt-syp}] 
\begin{align}
     \lim_{t\to\infty}\mathcal{W}/t&= \sigma_0 e^{-\alpha} (-\alpha)^{-b}\bigg[-(b+\alpha ) \Gamma (b)+\alpha  \Gamma (b,-\alpha )+\Gamma (b+1,-\alpha )\bigg]~+\nonumber\\
     &-\frac{e^{-\alpha} (-\alpha)^{-b} }{2 t_p}\bigg[-(b+\alpha ) (b+\alpha +1) \Gamma (b)+\alpha  (\alpha +1) \Gamma (b,-\alpha )~+\nonumber\\
    &+2 \alpha  \Gamma (b+1,-\alpha )+\Gamma (b+2,-\alpha )\bigg]\ , \label{scwk-lt}
\end{align}
{\color{black}where $\alpha\equiv 2 \sigma^2 A t_p(\theta + 2 t_p)$ and $b \equiv 2\sigma_0t_p - \alpha$ [both are defined below Eq.~\eqref{main:x2-t-val}]. Thus, the stationarity of the scaled average work requires $\sigma_0$ to be positive. (Notice that each term inside the square brackets on the right-hand side approaches to zero in the limit $\alpha\to 0$.) }

Next, we compute the average heat flow. The heat flow along a single stochastic trajectory~\eqref{ht-1]} can be rewritten using Eq.~\eqref{netru} as
\begin{align}
    q& = [U(x_t;k_t) - U(x_0;k_0)] - w\\
    &= \dfrac{\sigma_0}{2}[x_t^2 - x_0^2] +  \dfrac{\sigma}{2}\int_0^t~ds~k(s)\dot{V}\ . \label{q-eqn-1}
\end{align}
{\color{black}In the above Eq.~\eqref{q-eqn-1}, we average over an ensemble of trajectories $\{x\}_{0\leq t'\leq t}$  (for fixed initial condition $x_0=0$) for each fixed trajectory of $k(t)$, and subsequently substitute $\dot{\mathcal{V}}$ from Eq.~\eqref{var-eqn} on the right-hand side.}
Then, we perform average over an ensemble of trajectories of $\{k\}_{0\leq t'\leq t}$, and we obtain
\begin{align}
    \mathcal{Q} = \dfrac{\sigma_0}{2}\langle x^2\rangle -  \underbrace{\sigma\sigma_0\int_0^t~ds~\langle k(s)\mathcal{V}(s)\rangle_{\{k\}}}_{\mathcal{W}_1}-  \underbrace{\sigma^2\int_0^t~ds~\langle k^2(s)\mathcal{V}(s)\rangle_{\{k\}}}_{\mathcal{W}_2}\ , \label{ana-heat}
\end{align}
where $\langle x^2\rangle$ is given in Eq.~\eqref{main:x2-t-val}, and $\mathcal{W}_1$ and $\mathcal{W}_2$ are respectively evaluated in~\ref{si:w1} and \ref{si:w2}. In the long-time limit, $\langle x^2\rangle$ is independent of $t$~\eqref{x2-ltime}; therefore, $\lim_{t\to\infty} \mathcal{Q}/t$ is equal to the negative of $\lim_{t\to\infty} \mathcal{W}/t$~\eqref{scwk-lt}.

Figure~\ref{fig:avg-work} shows a good agreement of the analytical scaled average work~\eqref{lbo}, scaled average heat flow~\eqref{ana-heat}, their long-time expressions~\eqref{scwk-lt} and the negative of~Eq.~\eqref{scwk-lt}] with the numerical Langevin simulations.

The results in Fig.~\ref{fig:avg-work} can be interpreted as follows. Initially, the system starts from $x_0=0$, which corresponds to a non-stationary state of the system~\eqref{eqns}. [Notice that the stationary state of the system is characterized by the joint distribution $P_{ss}(x_0,k_0)$.] At this initial stage, the average work done is negligible, and the internal energy of the system increases due to heat absorption from the environment--resulting in a positive heat rate.

As time progresses, the stochastic switching of stiffness $k(t)$ leads to an increase in the average work done on the system. However, the system simultaneously begins to dissipate heat back into the environment. Consequently, the heat rate decreases while the work rate increases over time.

In the long-time limit, the system reaches a stationary state where the rate of change of average internal energy goes to zero. At this point, the average work performed on the system due to stiffness switching is exactly balanced by the rate of heat dissipation into the environment.

\section{Summary}
\label{sec:summ}
In summary, we investigated an 
overdamped Brownian particle in one dimension, that is confined in a harmonic trap with fluctuating stiffness. The stiffness 
was modeled via a stationary Ornstein–Uhlenbeck process with 
correlation time $t_p$, i.e., the particle in confined in a fluctuating stiffness harmonic trap. In the absence of the heat bath (deterministic limit, i.e., temperature $T=0$), we showed that the position's fluctuations are log-normal distributed. Moreover, we showed that this probability density function can be represented in the large-deviation form~\cite{Hugo-1}.

We investigated the position's fluctuations in the vanishing and long persistent time $t_p$ limit. For the vanishing $t_p$ case (i.e., while noise limit), the position distribution approaches a stationary power distribution with its tails decaying as $|x|^{-(1 + 2\sigma_0/a^2)}$ as $|x|\to \infty$, 
with $a$ being the parameter controlling the stiffness' fluctuation noise strength and $\sigma_0>0$. Herein, we showed the $n$th moment exist only if $a^2<2\sigma_0/n$. For the long persistent time $t_p$ (i.e., quenched noise limit), the exact analytical expression in the integral form is obtained. In the limit of vanishing quenched disorder, we obtained an analytical expression of the probability density function using the saddle-point approach. 

Further, for the case of finite non-zero $t_p$, we used the method of subordination~\cite{MOS-1,MOS-2} to exactly compute the $n$th position's moment and discussed the stationarity of these moments. It turned out that the moments exist only if $A\sigma^2(\theta + 2 t_p)<2\sigma_0/n$, which generalizes the condition of vanishing persistent time limit behavior.

From the thermodynamic point of view, this system is driven away from equilibrium by the stochastically fluctuating stiffness of the harmonic trap. To understand the energy flows associated with this system, we exactly computed the average work performed on the particle and the heat exchanged by the particle with the heat bath for all times as well as in the steady state.

We here remark that our results can be easily generalized 
to the case when $\sigma_0$ [see Eq.~\eqref{x-eqn-2}] is itself a random quenched disorder parameter, i.e., $\sigma_0\sim f(\sigma_0)$, 
with the normalized probability density function $f(\sigma_0)$. This can be done by performing 
a disorder average over $\sigma_0$, similarly as we did in Sec.~\ref{sec:QN-main} for stiffness, $k$, behaving as a quenched random variable.

Our study opens several research avenues for future directions. It would be great to extend our analysis to compute the thermodynamic properties of Brownian particle in a trap with fluctuating both its  minimum and stiffness (previously, researchers have investigated theoretically and experimentally the case of fluctuating trap's minimum~\cite{Gomez-Solano_2010, Apal-w-1}). Another extension would be to generalize the results of a many particle system as in Ref.~\cite{IRP-3} 
to the case of trap's stiffness 
{\color{black}fluctuating} according to a stationary Ornstein–Uhlenbeck process~[Eq.~\eqref{k-eqn} and \eqref{pk-ss}] or more generally to stiffness randomly fluctuating in an interval, $k(t)\in [k_-, k_+]$, with a constant rate. Further extension in the direction of many particle system would be to investigate the heat flow in harmonic chain~\cite{HF-1,HF-2} of fluctuating interactions. Finally, we emphasize that our results can be tested in an experiment by extending Ref.~\cite{goerlich2023experimental} to stochastically switching the intensity of laser beam confining the particle.

\ack
D.G. thanks Gregory Schehr and Jan Meibohm for many fruitful discussions. D.G. acknowledges the support from the Alexander von Humboldt foundation. \\
\bigskip

\appendix

\section{Derivation of Fokker-Planck equation for white noise stiffness}
\label{sec:FP-WN}
In general, the Fokker-Planck equation is given by~\cite{risken1989fokker}
\begin{align}
    \dfrac{\partial P(x,t|x_0)}{\partial t} =  -\dfrac{\partial}{\partial x} \bigg[\dfrac{\langle \Delta x\rangle}{\Delta t}\bigg|_{\Delta t \to 0} P(x,t|x_0)\bigg] +     \dfrac{1}{2}\dfrac{\partial^2}{\partial x^2} \bigg[\dfrac{\langle \Delta x^2\rangle}{\Delta t}\bigg|_{\Delta t \to 0} P(x,t|x_0)\bigg]\ . \label{gfpe}
\end{align}

For the white noise stiffness [Sec.~\ref{sec:WN-main} and Eq.~\eqref{corr-k-lim}], i.e., the temporal correlation of the stiffness are delta correlated, we have the following stochastic differential equation (SDE):
\begin{align}
    \dot{x}(t) = -(\sigma_0 + a \xi(t)]x(t) +\sqrt{2}\eta(t) \label{LE-WN}
\end{align}
where $\langle \eta(t)\eta(t')\rangle  = \delta(t-t')$ and $\langle \xi(t)\xi(t') \rangle =  \delta(t-t')$, and $a\equiv \sigma \sqrt{A\theta}$. 
We now aim to compute the Fokker-Planck equation corresponding to above SDE. To this end, we compute the average $\langle \Delta x(t) \rangle\equiv \langle x(t+\Delta) - x(t) \rangle/\Delta t$ and $\langle [\Delta x(t)]^2\rangle/\Delta t$ in the limit $\Delta t\to 0$.

We first discretize the Langevin dynamics~\eqref{LE-WN}
\begin{align}
    \Delta x(t)&= \sqrt{2}\int_t^{t+\Delta t}~ds~\eta(s) - \int_t^{t+\Delta t}~ds~[\sigma_0+a \xi(s)] x(s)\nonumber\\
    &=\sqrt{2}\int_t^{t+\Delta t}~ds~\eta(s) -x(\tau)\int_t^{t+\Delta t}~ds~[\sigma_0+a \xi(s)]\ ,
\end{align}
where we used the mean value theorem in the second integral for any $\tau \in [t,t+\Delta t]$. Now since $\xi(t)$ is a stochastic quantity, the choice of $\tau$ matters.  Let us say $x(\tau) \equiv \alpha x(t+\Delta t) + (1-\alpha)x(t) = x(t) + \alpha \Delta x(t)$ for any $\alpha$. Specifically, $\alpha=0, 1/2,1$, respectively, correspond to Ito, Stratonovich, and anti-Ito.

For a small $\Delta t$, we can approximate the above integrals, yielding 
\begin{align}
    \Delta x(t)   &=\underbrace{\sqrt{2}\int_t^{t+\Delta t}~ds~\eta(s)}_{I_0} - \underbrace{[x(t) + \alpha \Delta x(t)][\sigma_0\Delta t + a\int_t^{t+\Delta t}~ds~\xi(s)]}_{I_1}\ .
\end{align}

Let us now focus on the following integral:
\begin{align}
    I_1 &= [x(t) + \alpha \Delta x(t)][\sigma_0\Delta t + a\int_t^{t+\Delta t}~ds~\xi(s)]\ .\label{eq:I}
\end{align}

Substituting $\Delta x(t)$ from the above equation, we get
\begin{align}
    I_1 &= x(t)\bigg[\sigma_0\Delta t + a\int_t^{t+\Delta t}~ds~\xi(s)\bigg] + \alpha \bigg[ \sigma_0\Delta t + a\int_t^{t+\Delta t}~ds~\xi(s)\bigg] \bigg[\sqrt{2}\int_t^{t+\Delta t}~ds'~\eta(s') \nonumber\\
    &-[x(t)+\alpha \Delta x(t)][\sigma_0\Delta t +a\int_t^{t+\Delta t}~ds'~\xi(s')\bigg]
\end{align}
Averaging over noises, we get
\begin{align}
    \langle I_1 \rangle &= \Delta t \sigma_0 x(t) -\bigg\langle \alpha x(t) [\sigma_0\Delta t + a \int_t^{t+\Delta t}~ds~\xi(s)][\sigma_0\Delta t + a\int_t^{t+\Delta t}~ds'~\xi(s')]\bigg\rangle\nonumber\\
    &= \Delta t [ \sigma_0 - \alpha a^2] x(t) \, \label{eq:avg-I}
\end{align}
where we dropped terms higher order than $\Delta t$ including $\Delta x$ (which will eventually gives terms higher order than $\Delta t$). This implies 
\begin{align}
    \langle  \Delta x(t) \rangle &= -\Delta t [ \sigma_0 - \alpha a^2] x(t)
\end{align}
yielding 
the first term in the Fokker-Planck equation: $\lim_{\Delta t\to 0}\langle \Delta x\rangle/\Delta t$.

Proceeding similarly as above, we compute the second moment:
\begin{align}
    \langle [\Delta x(t)]^2\rangle &= \bigg\langle 2\int_t^{t+\Delta t}~ds'~\eta(s)\eta(s') + [x(t)+\alpha \Delta x(t)]^2 [\sigma_0^2\Delta t^2 +a^2 \int_t^{t+\Delta t}~ds'~\xi(s)\xi(s')\nonumber\\
    &+ 2 \Delta t \sigma_0 a \int_t^{t+\Delta t}~ds~\xi(s)]\bigg\rangle \ .
\end{align}
From this, we obtain
\begin{align}
        \langle [\Delta x(t)]^2\rangle = [2 + x(t)^2 a^2 ]\Delta t
\end{align}
which gives the second term in the Fokker-Planck equation: $\lim_{\Delta t\to 0}\langle \Delta x^2\rangle/\Delta t$.

Using Eq.~\eqref{gfpe}, the Fokker-Planck equation corresponding to SDE~\eqref{LE-WN} results as 
\begin{align}
    \dfrac{\partial P(x,t|x_0)}{\partial t} =  \dfrac{\partial}{\partial x} \bigg[(\sigma_0 - \alpha a^2) x P(x,t|x_0)\bigg] +     \dfrac{\partial^2}{\partial x^2} \bigg[ (1 + a^2 x^2/2) P(x,t|x_0)\bigg]\ ,
\end{align}
where the second term (proportional to $a^2$) in each of the square brackets on the right-hand side is due to the white noise stiffness.

\section{Evaluation of integrals $\mathcal{I}_1$~\eqref{def-I1} and $\mathcal{I}_1$~\eqref{def-I2}}
\label{sec:I1I2}
\subsection{Calculation of $\mathcal{I}_1$}
\label{sec:I1}
\begin{align}
   \mathcal{I}_1 &= \int_0^t~ds~ e^{-2\sigma_0s}~e^{-2\sigma^2 H_1(s)}\nonumber\\
   &=e^{-\alpha}\int_0^t~ds~ e^{-s/t_p(2\sigma_0 t_p -\alpha)}~e^{\alpha e^{-s/t_p}}
\end{align}
where we defined 
\begin{align}
    \alpha \equiv 2 \sigma^2 A t_p(\theta + 2 t_p)\ .
\end{align}

Substituting $z=e^{-s/t_p}$, we get 
\begin{align}
   \mathcal{I}_1 &= t_p  e^{-\alpha } \int_{e^{-t/t_p}}^1~dz~z^{a-1}e^{\alpha z}\nonumber\\
   &= t_p  e^{-\alpha }(-\alpha )^{-a} \left[\Gamma \left(a,-e^{-\frac{t}{t_p}} \alpha \right)-\Gamma (a,-\alpha )\right]\ , \label{x2-t-val}
\end{align}
where $\Gamma(a,z) = \int_z^\infty~x^{a-1}e^{-x}$ is the incomplete gamma function. The integral is evaluated for $t_p>0$ and $a = 2\sigma_0 t_p - \alpha$.
In the limit $t\to \infty$, $\Gamma \left(a,-e^{-\frac{t}{t_p}} \alpha \right)$ becomes $\Gamma (a)$ for $a>0$.
\begin{align}
   \mathcal{I}_1 
   &= t_p  e^{-\alpha }(-\alpha )^{-a} \left[\Gamma \left(a \right)-\Gamma (a,-\alpha )\right]\ ,
\end{align}

\subsection{$\mathcal{I}_2$}
\label{sec:I2}
\begin{align}
  \mathcal{I}_2 &=  \int_0^t~ds e^{2\sigma_0 s} e^{-2\sigma^2[H_2(s,s)  -3H_2(t,s)]}\ .
\end{align}
Substituting $z = e^{-s/t_p}$, we get
\begin{align}
    \mathcal{I}_2 &=t_p~e^{\alpha/2} e^{-\frac{3 \alpha}{2} e^{-t/t_p}}  \int_{e^{-t/t_p}}^1~dz~z^{2 \alpha -2\sigma_0tp -1} e^{-\alpha  z/2} e^{\frac{3 \alpha}{2} \frac{ e^{-t/t_p}}{z}}\\
     &=t_p~e^{\alpha/2} e^{-\frac{3 \alpha}{2}  e^{-t/t_p}}  \int_{e^{-t/t_p}}^1~dz~z^{\alpha -a -1} e^{-\alpha  z/2} e^{\frac{3 \alpha}{2}\frac{e^{-t/t_p}}{z}}\\
     &=t_p~e^{\alpha/2} e^{-\frac{3 \alpha}{2}  e^{-t/t_p}}  \sum_{m=0}^\infty \dfrac{(3\alpha/2)^m e^{-mt/t_p} }{m!}\int_{e^{-t/t_p}}^1~dz~z^{\alpha -a -1-m} e^{-\alpha  z/2} \\
     & =t_p~e^{\alpha/2} e^{-\frac{3 \alpha}{2}  e^{-t/t_p}}  \sum_{m=0}^\infty \dfrac{(3\alpha/2)^m e^{-mt/t_p} }{m!} \left[e^{\frac{t (a-\alpha +m)}{t_p}} E_{a+m-\alpha +1}\left(\frac{\alpha e^{-\frac{t}{t_p}}}{2} \right)-E_{a+m-\alpha +1}\left(\frac{\alpha }{2}\right)\right]\ ,
\end{align}
where $E_n(y)=\int_1^{\infty }~dx~x^{-n} e^{-xy}$ is the exponential integral function.

\section{Calculations of average $\mathcal{W}$}
\label{app:work}
\begin{align}
    \mathcal{W}= \underbrace{\dfrac{\sigma}{2}\langle k(t) \mathcal{V}(t) \rangle_{\{k\}}}_{\mathcal{W}_0} 
+ \underbrace{\sigma \sigma_0 \int_0^t~ds~\langle k(s) {\mathcal{V} (s)}\rangle_{\{k\}}}_{\mathcal{W}_1}  +\underbrace{\sigma^2 \int_0^t~ds~\langle k^2(s) {\mathcal{V} (s)}\rangle_{\{k\}}}_{\mathcal{W}_2}  \ , \label{si:lbo}
\end{align}
To compute this average work, we have to compute the following annealed averages, i.e., averages over the stochastic trajectories of $k(t)$
\begin{align}
    A_n = \langle k^n(u)\mathcal{V}(u) \rangle_{\{k(t)\}}\ , \label{A-n-eqn}
\end{align}
for $n = 1,2$. For convenience, in the following, we drop the subscript ${\{k(t)\}}$.

We aim to compute the average work for a fixed initial condition $x_0=0$.
The second moment for a given realization of $k(t)$ is the solution of Eq.~\eqref{sec-mom-v-eqn}
\begin{align}
    \mathcal{V} (u)= 2\int_0^u~ds~e^{-2\sigma_0(u-s)}e^{-2\sigma\int_{s}^u~ds'~k(s')}\ .
\end{align}

Then, the average for $n=1$ in Eq.~\eqref{A-n-eqn} is 
\begin{align}
   \langle k(u) {\mathcal{V} (u)}\rangle = 2 \int_0^u~ds~e^{-2\sigma_0(u-s)}\bigg\langle k(u)e^{-2\sigma\int_{s}^u~ds'~k(s')}\bigg\rangle  \ . \label{k1vu-eqn}
\end{align}

Since the average of $k(t)$ in the stationary-state~\eqref{pk-ss} is zero, the non-zero contributions arise from the odd-multiple terms of $k(t)$ from the expansion of the exponential  
\begin{align}
   &\bigg\langle k(u)e^{-2\sigma\int_{s}^u~ds'~k(s')}\bigg\rangle = \bigg\langle k(u)\bigg[ -2\sigma \int_{s}^u~dm_1~k(m_1) ~+ \nonumber\\
   &- \dfrac{(2\sigma)^3}{3!}\int_{s}^u~dm_1~\int_{s}^u~dm_2~\int_{s}^u~dm_3~k(m_1)k(m_2)k(m_3)~+\nonumber\\
   &- \dfrac{(2\sigma)^5}{5!}\int_{s}^u~dm_1~\int_{s}^u~dm_2~\int_{s}^u~dm_3~\int_{s}^u~dm_4~\int_{s}^u~dm_5~k(m_1)k(m_2)k(m_3)k(m_4)k(m_5)+\dots \bigg]\bigg\rangle\nonumber\\
   &\equiv T^{(1)}_1 (s,u) + T^{(1)}_2(s,u) + T^{(1)}_3(s,u)\ , \label{kv-eq1}
\end{align}
where
\begin{align}
   T^{(1)}_1 &=   -2\sigma \int_{s}^u~dm_1~\langle k(u)k(m_1)\rangle\ , \\
   T^{(1)}_2 &=  - \dfrac{(2\sigma)^3}{3!}\int_{s}^u~dm_1~\int_{s}^u~dm_2~\int_{s}^u~dm_3~\langle k(t)k(m_1)k(m_2)k(m_3)\rangle \nonumber\\
   &= - 3\dfrac{(2\sigma)^3}{3!}\int_{s}^u~dm_1~\int_{s}^u~dm_2~\int_{s}^u~dm_3~\langle k(t)k(m_1)\rangle \langle k(m_2)k(m_3)\rangle\nonumber\\
   &=T^{(1)}_1 \dfrac{(2\sigma)^2}{2}\int_{s}^u~dm_1~\int_{s}^u~dm_2~\langle k(m_1) k(m_2)\rangle\ ,\\
   T^{(1)}_3 &= - \dfrac{(2\sigma)^5}{5!}\int_{s}^u~dm_1~\int_{s}^u~dm_2~\int_{s}^u~dm_3~\int_{s}^u~dm_4~\int_{s}^u~dm_5\langle k(t)k(m_1)k(m_2)k(m_3)k(m_4)k(m_5)\rangle\nonumber\\
   &=-15\dfrac{(2\sigma)^5}{5!}\int_{s}^u~dm_1~\int_{s}^u~dm_2~\int_{s}^u~dm_3~\int_{s}^u~dm_4~\int_{s}^u~dm_5\langle k(t)k(m_1)\rangle \langle k(m_2)k(m_3)\rangle \langle k(m_4)k(m_5)\rangle\nonumber\\
   & = T^{(1)}_1 \dfrac{\sigma^2}{2!}\int_{t'}^t~dm_1~\int_{s}^u~dm_2~\int_{s}^u~dm_3~\int_{s}^u~dm_4\langle k(m_2)k(m_3)\rangle \langle k(m_4)k(m_5)\rangle\ ,
\end{align}
where we used the Wick's theorem to simplify the higher-point correlations into two-point correlations. Thus, Eq.~\eqref{k1vu-eqn} becomes
\begin{align}
   \langle k(u) {\mathcal{V} (u)}\rangle &
   =2 \int_0^u~ds~e^{-2\sigma_0(u-s)} \sum_{i=1}^3~T_i^{(1)}(s_1,u)\nonumber\\
   &=2 \int_0^u~ds~e^{-2\sigma_0(u-s)}~T_1^{(1)}(s,u) e^{2\sigma^2\int_{s}^u~dm_1~\int_{s}^u~dm_2~\langle k(m_1) k(m_2)\rangle}\nonumber\\
   &=-4\sigma \int_0^u~ds~~H_3(u-s)~e^{-2\sigma_0(u-s)}e^{-2\sigma^2 H_1(u-s)}\nonumber\\
   &=-4\sigma \int_0^u~ds~~H_3(u)~e^{-2\sigma_0u}e^{-2\sigma^2 H_1(u)}\ , \label{ku-app-1}
\end{align}
where we defined 
\begin{align}
    H_1(u)&\equiv At_p(\theta+2t_p)(1 - u/t_p -e^{-u/t_p})\ , \label{h1u-2}\\
    H_3(u)&\equiv  \frac{1}{2} A (\theta +2 t_p) \left(1-e^{-\frac{u}{t_p}}\right)\ . \label{h3u-1}
\end{align}

Let us now compute the average for $n=2$ in Eq.~\eqref{A-n-eqn}
\begin{align}
   \langle k^2(u) {\mathcal{V} (u)}\rangle &= 2 \int_0^u~ds~e^{-2\sigma_0(u-s)}\bigg\langle k^2(u)e^{-2\sigma\int_{s}^u~ds'~k(s')}\bigg\rangle\ . \label{k2vu0-eqn}
\end{align}
Similar to the calculation for $n=1$, here the even multiple of $k(t)$ terms from the exponential contribute; therefore, 
\begin{align}
  \bigg\langle &k^2(u)e^{-2\sigma\int_{s}^u~ds'~k(s')}\bigg\rangle = \bigg\langle k^2(u)\bigg[1 + \dfrac{(2\sigma)^2}{2!}\int_{s_1}^u~dm_1~\int_{s_1}^u~dm_2~k(m_1)k(m_2)+\nonumber\\
  & + \dfrac{(2\sigma)^4}{4!}\int_{s_1}^u~dm_1~\int_{s_1}^u~dm_2~\int_{s_1}^u~dm_3~\int_{s_1}^u~dm_4~k(m_1)k(m_2)k(m_3)k(m_4)+ \dots\bigg]\bigg\rangle\nonumber\\
   &\equiv  T_1^{(2)}(s,u) +  T_2^{(2)}(s,u) +  T_3^{(2)}(s,u)\ ,
\end{align}
where
\begin{align}
    T_1^{(2)} &= \langle k^2(u) \rangle\ ,\\
    T_2^{(2)} &= \dfrac{(2\sigma)^2}{2!}\int_{s}^u~dm_1~\int_{s}^u~dm_2~\langle k^2(u)k(m_1)k(m_2\rangle \nonumber\\
    & = T_1^{(2)} \dfrac{(2\sigma)^2}{2}\int_{s}^u~dm_1~\int_{s}^u~dm_2~\langle k(m_1)k(m_2)\rangle + (2\sigma)^2\bigg[\int_{s}^u~ds~\langle k(u)k(s)\rangle \bigg]^2\ ,\\
    T_3^{(2)}&=3T_1^{(2)} \dfrac{(2\sigma)^4}{4!} \bigg[\int_{s}^u~dm_1~\int_{s_1}^u~dm_2~\langle k(m_1)k(m_2)\rangle\bigg]^2 + 12 \dfrac{(2\sigma)^4}{4!} \bigg[\int_{s}^u~dm_1~\langle k(u)k(m_1)\rangle\bigg]^2~\times \nonumber\\
    &\times \bigg[\int_{s}^u~dm_3~\int_{s}^u~dm_4~\langle k(m_3)k(m_4)\rangle\bigg]\nonumber\\
    &=T_1^{(2)} \dfrac{((2\sigma)^2/2)^2}{2!} \bigg[\int_{s}^u~dm_1~\int_{s}^u~dm_2~\langle k(m_1)k(m_2)\rangle\bigg]^2 + \dfrac{(2\sigma)^4}{2}\bigg[\int_{s}^u~dm_1~\langle k(u)k(m_1)\rangle\bigg]^2~\times \nonumber\\
    &\times\bigg[\int_{s}^u~dm_3~\int_{s}^u~dm_4~\langle k(m_3)k(m_4)\rangle\bigg]\ .
\end{align}

Therefore, Eq.~\eqref{k2vu0-eqn} becomes
\begin{align}
\langle k^2(u) {\mathcal{V} (u)}\rangle &= 2 \int_0^u~ds~e^{-2\sigma_0(u-s)}\sum_{i=1}^3T_i^{(2)}(s,u)\nonumber\\
&=2 \int_0^u~ds~e^{-2\sigma_0(u-s_1)}\bigg[T_1^{(2)} + 4\sigma^2\bigg(\int_{s}^u~dm_1~\langle k(u)k(m_1)\rangle\bigg)^2 \bigg]~\times \nonumber\\
&\times \exp\bigg[2\sigma^2\int_{s}^u~dm_1~\int_{s}^u~dm_2~\langle k(m_1) k(m_2)\rangle\bigg]\nonumber\\
&=2 \int_0^u~ds~\bigg[A \dfrac{(\theta+2 t_p)}{2t_p} +4 \sigma^2 H_3^2(u-s) \bigg] e^{-2\sigma_0(u-s)}e^{-2\sigma^2 H_1(u-s)}\nonumber\\
&=2 \int_0^u~ds~\bigg[A \dfrac{(\theta+2 t_p)}{2t_p} +4 \sigma^2 H_3^2(u) \bigg] e^{-2\sigma_0u}e^{-2\sigma^2 H_1(u)\rangle}\label{si-k2u}\ ,
\end{align}
where $H_1(u)$ and $H_3(u)$ are defined in Eqs.~\eqref{h1u-2} and \eqref{h3u-1}.

\subsection{Calculations of $\mathcal{W}_{0,1,2}$~\eqref{si:lbo}}
\label{app:w0w1w2}
In the following, we calculate the integrals involving in $\mathcal{W}_{0,1,2}$~\eqref{si:lbo}. 
\subsubsection{Calculation of $\mathcal{W}_0$:}
\label{si:w0}
Here, we calculate $\mathcal{W}_0$, where we need to compute the following average [see Eq.~\eqref{ku-app-1}]
\begin{align}
   \langle k(s) {\mathcal{V} (s)}\rangle_{\{k\}} &=-4\sigma \int_0^s~du~~H_3(u)~e^{-2\sigma_0 u}e^{-2\sigma^2 H_1(u)}\ , \label{ku}
\end{align}
where $H_1(u)$ and $H_3(u)$ are respectively given in Eqs.~\eqref{h1u-2} and~\eqref{h3u-1}.

We simplify the above integral~\eqref{ku}
\begin{align}
   \langle k(s) {\mathcal{V} (s)}\rangle_{\{k\}} 
    &=-\frac{\alpha}{\sigma t_p} e^{-\alpha} \int_0^s~du~\left(e^{- u/t_p(2\sigma_0 t_p - \alpha)}-e^{-\frac{u}{t_p}(2\sigma_0 t_p+1 - \alpha)}\right) e^{\alpha e^{-u/t_p}}\ , \label{ku-check}
\end{align}
where, for convenience, we defined $\alpha \equiv 2 \sigma^2 A t_p(\theta + 2 t_p)$.

Substituting $e^{-u/t_p}=z$, we get
\begin{align}
   \langle k(s) {\mathcal{V} (s)}\rangle_{\{k\}} 
    &=\frac{\alpha}{\sigma} e^{-\alpha} \int_1^{e^{-s/t_p}}~dz~\left[z^{(2\sigma_0t_p - \alpha - 1)}-z^{(2\sigma_0t_p - \alpha)}\right] e^{\alpha z}\nonumber\\
    & = \frac{e^{-\alpha} (-\alpha)^{-b}}{\sigma }\left[\alpha \Gamma (b,-\alpha)+\Gamma (b+1,-\alpha)-\alpha \Gamma \left(b,-e^{-\frac{s}{t_p}} \alpha\right)-\Gamma \left(b+1,-e^{-\frac{s}{t_p}} \alpha\right)\right]\ , \label{ku-exp2}
\end{align}
where $\Gamma(b,z) = \int _z^{\infty }dx~ x^{b-1} e^{-x}$ is the incomplete gamma function, and we again defined
\begin{align}
    b\equiv 2 \sigma_0 t_p - \alpha\ .
\end{align}

The right-hand side of Eq.~\eqref{ku-exp2} in the long-time limit converges for $b>0$:
\begin{align}
    \lim_{t\to \infty}\langle k(t) {\mathcal{V} (t)}\rangle_{\{k\}} \to \frac{e^{-\alpha} (-\alpha)^{-b}}{\sigma } [-(b+\alpha ) \Gamma (b)+\alpha  \Gamma (b,-\alpha )+\Gamma (b+1,-\alpha )]\ . \label{kulongt}
\end{align}

Therefore, the contribution from the scaled average $\mathcal{W}_0/t$ in the long time limit is:
\begin{align}
    \lim_{t\to \infty}\mathcal{W}_0/t = 0\ . \label{w0bt_asyp}
\end{align}

\subsubsection{Calculation of $\mathcal{W}_1$:}
\label{si:w1}
To evaluate the second term in $\mathcal{W}_1$ in Eq.~\eqref{si:lbo} we see that
\begin{align}
    \mathcal{W}_1 &= \sigma \sigma_0 \int_0^t~ds~\langle k(s) {\mathcal{V} (s)}\rangle_{\{k\}} \nonumber\\
    &= -\frac{\alpha \sigma_0}{t_p} e^{-\alpha} \int_0^t~ds~\int_0^s~du~\left(e^{-2\sigma_0 u}-e^{-\frac{u}{t_p}(2\sigma_0t_p+1)}\right)e^{\alpha u/t_p} e^{\alpha e^{-u/t_p}}\ , \label{w1-first}
\end{align}
where we used Eq.~\eqref{ku-check}.

Swapping the order of integration on the right-hand side of Eq.~\eqref{w1-first}, we get
\begin{align}
    \mathcal{W}_1 &= -\frac{\alpha \sigma_0}{t_p} e^{-\alpha} \int_0^t~du~(t-u)~\left(e^{-2\sigma_0 u}-e^{-\frac{u}{t_p}(2\sigma_0t_p+1)}\right)e^{\alpha u/t_p} e^{\alpha e^{-u/t_p}}\\
    &=\underbrace{\sigma \sigma_0 t \langle k(t) {\mathcal{V} (t)}\rangle_{\{k\}}}_{\mathcal{W}_1^{(1)}} +\underbrace{\frac{\alpha \sigma_0}{t_p} e^{-\alpha} \int_0^t~du~u~\left(e^{-2\sigma_0 u}-e^{-\frac{u}{t_p}(2\sigma_0t_p+1)}\right)e^{\alpha u/t_p} e^{\alpha e^{-u/t_p}}}_{\mathcal{W}_1^{(2)}}\ .\label{i4-eqn}
\end{align}
We evaluate $\mathcal{W}_1^{(1)}$ at time $t$ using Eq.~\eqref{ku-exp2}, while we substitute $z = e^{-u/t_p}$ in $\mathcal{W}_1^{(2)}$~\eqref{i4-eqn}; this gives
\begin{align}
    \mathcal{W}_1^{(2)}
    &=\alpha \sigma_0 t_p  e^{-\alpha} \int_1^{e^{-t/t_p}}~dz~\ln z~\left[z^{(2\sigma_0t_p - \alpha - 1)}-z^{(2\sigma_0t_p - \alpha)}\right] e^{\alpha z}\nonumber\\
    &=e^{-\alpha } \sigma_0 (-\alpha )^{-b} \bigg[t_p \bigg\{\alpha  G_{2,3}^{3,0}\bigg(-\alpha \left|
\begin{array}{c}
 1,1 \\
 0,0,b \\
\end{array}
\right.\bigg)+G_{2,3}^{3,0}\left(-\alpha \left|
\begin{array}{c}
 1,1 \\
 0,0,b+1 \\
\end{array}
\right.\right)~ +\nonumber\\
&-\alpha  G_{2,3}^{3,0}\left(-e^{-\frac{t}{t_p}} \alpha \bigg|
\begin{array}{c}
 1,1 \\
 0,0,b \\
\end{array}
\right)-G_{2,3}^{3,0}\left(-e^{-\frac{t}{t_p}} \alpha \bigg|
\begin{array}{c}
 1,1 \\
 0,0,b+1 \\
\end{array}
\right)\bigg\}~+\nonumber\\
&+t \bigg(\alpha  \Gamma \left(b,-e^{-\frac{t}{t_p}} \alpha \right)+\Gamma \left(b+1,-e^{-\frac{t}{t_p}} \alpha \right)\bigg)\bigg]\ , \label{i4-fin-1}
\end{align}
where $G(...)$ is the Meijer G function.

Further, the integral $\mathcal{W}_1^{(2)}$~\eqref{i4-fin-1} in the long-time limit converges (and becomes independent of $t$) for $b>0$, and this gives
\begin{align}
    \lim_{t\to\infty} \mathcal{W}_1^{(2)} = e^{-\alpha } \alpha  \sigma_0 t_p \left(\frac{\, _2F_2(b,b;b+1,b+1;\alpha )}{b^2}-\frac{\, _2F_2(b+1,b+1;b+2,b+2;\alpha )}{(b+1)^2}\right)\ ,
\end{align}
where $_pF_q(b_1;b_2;z)$ is the generalized hypergeometric function.

Thus, from Eq.~\eqref{i4-eqn}, we see that the dominant contribution to $\mathcal{W}_1/t$ in the long-time limit comes from $\mathcal{W}_1^{(1)}$ (for $b>0$), yielding
\begin{align}
    \lim_{t\to\infty}\mathcal{W}_1/t= 
    \sigma_0 e^{-\alpha} (-\alpha)^{-b}[-(b+\alpha ) \Gamma (b)+\alpha  \Gamma (b,-\alpha )+\Gamma (b+1,-\alpha )] \label{w1bt-syp}
\end{align}

\subsubsection{Calculations of $\mathcal{W}_2$}
\label{si:w2}
\begin{align}
    \mathcal{W}_2&= \sigma^2 \int_0^t~ds~\langle k^2(s) {\mathcal{V} (s)}\rangle_{\{k\}}\label{s1-w2-f}\\
    &=\mathcal{W}_2^{(1)} + \mathcal{W}_2^{(2)}\ , \label{app:W_2}
\end{align}
where we substituted the expression~\eqref{si-k2u} in Eq.~\eqref{s1-w2-f}, and then, substitute $z=e^{-s/t_p}$. This leads to the following integrals
\begin{align}
    \mathcal{W}_2^{(1)} \equiv -\dfrac{\alpha}{2 t_p}e^{-\alpha}t \int_0^{e^{-t/t_p}}~dz~[(1+\alpha)z^{b-1} + \alpha(z^{b+1}-2z^{b})]e^{\alpha z}\ , \\
        \mathcal{W}_2^{(2)} \equiv -\dfrac{\alpha}{2}e^{-\alpha} \int_0^{e^{-t/t_p}}dz~\ln z[(1+\alpha)z^{b-1} + \alpha(z^{b+1}-2z^{b})]e^{\alpha z}\ ,
\end{align}
where $\alpha$ is defined below Eq.~\eqref{ku-check}.

Computing $\mathcal{W}_2^{(1)}$ and $\mathcal{W}_2^{(2)}$, we get
\begin{align}
    \mathcal{W}_2^{(1)}&=-\frac{t e^{-\alpha} (-\alpha)^{-b} }{2 t_p}\bigg[\Gamma (b+2,-\alpha )+\alpha  \bigg\{(\alpha +1) \Gamma (b,-\alpha )+2 \Gamma (b+1,-\alpha )~+\nonumber\\
    &-(\alpha +1) \Gamma \left(b,-e^{-\frac{t}{t_p}} \alpha \right)-2 \Gamma \left(b+1,-e^{-\frac{t}{t_p}} \alpha \right)\bigg\}-\Gamma \left(b+2,-e^{-\frac{t}{t_p}} \alpha \right)\bigg]\ ,\\
    \mathcal{W}_2^{(2)}&=-\frac{e^{-\alpha } (-\alpha )^{-b} }{2 t_p}
    \bigg[-\alpha ^2 t_p G_{2,3}^{3,0}\left(-e^{-\frac{t}{t_p}} \alpha |
\begin{array}{c}
 1,1 \\
 0,0,b \\
\end{array}
\right)-\alpha  t_p G_{2,3}^{3,0}\left(-e^{-\frac{t}{t_p}} \alpha |
\begin{array}{c}
 1,1 \\
 0,0,b \\
\end{array}
\right)~\nonumber\\
&-2 \alpha  t_p G_{2,3}^{3,0}\left(-e^{-\frac{t}{t_p}} \alpha |
\begin{array}{c}
 1,1 \\
 0,0,b+1 \\
\end{array}
\right)-t_p G_{2,3}^{3,0}\left(-e^{-\frac{t}{t_p}} \alpha |
\begin{array}{c}
 1,1 \\
 0,0,b+2 \\
\end{array}
\right)~+\nonumber\\
&-(\alpha +1) \alpha ^2 t_p G_{2,3}^{3,0}\left(-\alpha \left|
\begin{array}{c}
 0,0 \\
 -1,-1,b-1 \\
\end{array}
\right.\right)-2 \alpha ^2 t_p G_{2,3}^{3,0}\left(-\alpha \left|
\begin{array}{c}
 0,0 \\
 -1,-1,b \\
\end{array}
\right.\right)~+\nonumber\\
&-\alpha  t_p G_{2,3}^{3,0}\left(-\alpha \left|
\begin{array}{c}
 0,0 \\
 -1,-1,b+1 \\
\end{array}
\right.\right)+\alpha ^2 t \Gamma \left(b,-e^{-\frac{t}{t_p}} \alpha \right)+\alpha  t \Gamma \left(b,-e^{-\frac{t}{t_p}} \alpha \right)~+\nonumber\\
&+2 \alpha  t \Gamma \left(b+1,-e^{-\frac{t}{t_p}} \alpha \right)+t \Gamma \left(b+2,-e^{-\frac{t}{t_p}} \alpha \right)\bigg]\ .
\end{align}
Substituting these $\mathcal{W}_2^{(1),(2)}$ in Eq.~\eqref{app:W_2} gives $\mathcal{W}_2$.

In the long-time limit, both $\mathcal{W}_2^{(1)}$ and $\mathcal{W}_2^{(2)}$ converge and behave as (for $b>0$)
\begin{align}
    \mathcal{W}_2^{(1)}&=-\frac{t e^{-\alpha} (-\alpha)^{-b} }{2 t_p}[-(b+\alpha ) (b+\alpha +1) \Gamma (b)+\alpha  (\alpha +1) \Gamma (b,-\alpha )~+\nonumber\\
    &+2 \alpha  \Gamma (b+1,-\alpha )+\Gamma (b+2,-\alpha )]\label{eqw21}\ ,\\
    \mathcal{W}_2^{(2)}&=-\frac{1}{2} e^{-\alpha } \alpha\bigg[\alpha  \Gamma (b+2)^2 \bigg(\, _2\tilde{F}_2(b+2,b+2;b+3,b+3;\alpha )~+\nonumber\\
    &-\frac{2 \, _2\tilde{F}_2(b+1,b+1;b+2,b+2;\alpha )}{(b+1)^2}\bigg)+\frac{(\alpha +1) \, _2F_2(b,b;b+1,b+1;\alpha )}{b^2}\bigg]\ .
\end{align}

Therefore, in the long-time limit, the dominant contribution to $\mathcal{W}_2/t$  comes from $\mathcal{W}_2^{(1)}$~\eqref{eqw21}
\begin{align}
    \lim_{t\to\infty}\mathcal{W}_2/t &= -\frac{e^{-\alpha} (-\alpha)^{-b} }{2 t_p}[-(b+\alpha ) (b+\alpha +1) \Gamma (b)+\alpha  (\alpha +1) \Gamma (b,-\alpha )~+\nonumber\\
    &+2 \alpha  \Gamma (b+1,-\alpha )+\Gamma (b+2,-\alpha )]\ . \label{w2bt-syp}
\end{align}

\section*{References}
\providecommand{\newblock}{}

\end{document}